\documentclass{JHEP3} 
\usepackage{epsfig}
\def\cA{{\cal A}}

\def\nn{\nonumber}

\def\la{\langle}
\def\ra{\rangle}

\def\l{\left}
\def\r{\right}

\def\beq{\begin{equation}}
\def\eeq{\end{equation}}
\def\bea{\begin{eqnarray}}
\def\eea{\end{eqnarray}}
\def\barr{\begin{array}}
\def\earr{\end{array}}
\def\be{\begin{equation}}
\def\ee{\end{equation}}
\def\bc{\begin{center}}
\def\ec{\end{center}}
\def\dd{\displaystyle}
\def\ov{\overline}

\def\Ne#1{{\cal N}=#1}
\def\De#1{{\it D}=#1}

\def\de{\partial}

\def\Im{\mathop{\mbox{Im}}}
\def\Re{\mathop{\mbox{Re}}}

\title{D~terms from D-branes, gauge invariance and \\ 
moduli stabilization in flux compactifications} 
\author{
Giovanni Villadoro \\ 
Jefferson Physical Laboratory, Harvard University, \\
Cambridge, Massachusetts 02138, USA \\
E-mail: \email{villador@physics.harvard.edu}}
\author{
Fabio Zwirner \\
Dipartimento di Fisica, Universit\`a di Padova and INFN, \\ 
Sezione di Padova, Via Marzolo 8, I-35131 Padova, Italy \\
E-mail: \email{fabio.zwirner@pd.infn.it}}
\preprint{DFPD-06/TH/01 \\ HUTP-06/A0003}
\abstract{ We elucidate the structure of D~terms in $\Ne1$ orientifold 
compactifications with fluxes. As a case study, we consider a simple orbifold 
of the type-IIA theory with D6-branes at angles, O6-planes and general NSNS, RR and 
Scherk-Schwarz geometrical fluxes. We examine in detail the emergence of D~terms, in
their standard supergravity form, from an appropriate limit of the D-brane action. 
We derive the consistency conditions on gauged symmetries and 
general fluxes coming from brane-localized Bianchi identities, and their relation with the 
Freed-Witten anomaly. We extend our results to other $\Ne1$
compactifications and to non-geometrical fluxes.  Finally, we discuss
the possible r\^ole of $U(1)$ D~terms in the stabilization of the untwisted moduli from the 
closed string sector. 
}
\keywords{Compactification and String Models, Supersymmetry Breaking, Field
Theories in Higher Dimensions, Supergravity Models}
\begin{document} 
%
%%%%%%%%%%%%%%%%%%%%%%%%%%%%%%%%%%%%%%%
\section{Introduction}
Orientifold compactifications of type-II superstrings with branes and fluxes (for some 
recent reviews and references, see e.g. \cite{IIrev, uranga, bcls}) offer new possibilities 
not only for obtaining the Standard Model spectrum and renormalizable interactions, but 
also for addressing long-standing problems such as moduli stabilization and supersymmetry 
breaking. We focus here on the last two problems. The ultimate goal is to find a 
string vacuum with spontaneously broken supersymmetry on an approximately flat 
four-dimensional background, with tiny positive vacuum energy density and all moduli 
stabilized. Eventually, we would also like to understand how `our' vacuum may be selected 
among other possible ones. We are still far from this ambitious goal, but important progress 
is being made.

The phenomenologically relevant constructions under better theoretical control
are those in which the effective potential for the light modes is generated
at  the classical level by fluxes, and can be described by an effective $\Ne1$, 
$D=4$ supergravity, obtained from the underlying higher-dimensional theory 
via generalized dimensional reduction. It may well be that perturbative and
non-perturbative quantum corrections play a crucial r\^ole in the complete 
resolution of the problems, but it is interesting to explore how far we can go at 
this classical level. 

A particularly relevant set of fields, whose dynamics is crucial for the 
above-mentioned problems, are the closed string moduli associated 
with the spin-0 fluctuations of the dilaton, of the metric and of the $p$-form 
potentials in the Neveu-Schwarz Neveu-Schwarz (NSNS) and Ramond-Ramond
(RR) sectors. In $\Ne1$ compactifications, these moduli can be assigned to chiral 
supermultiplets. For the sake of illustration, we concentrate here on the $T^6 / (Z_2 
\times Z_2)$ orbifold supplemented by a suitable orientifold, but  our results 
apply also to other $\Ne1$ compactifications. In the case under consideration, 
there are seven `main' moduli, denoted by
\beq
S = s + i \sigma \, , 
\qquad 
T_A = t_A + i \tau_A \, , 
\qquad 
U_A = u_A + i \nu_A \, ,
\label{studef}
\eeq
where $A=1,2,3$ corresponds to the three factorized 2-tori defined by the orbifold.
Here and in the following, we call {\it main} moduli the seven complex scalars in 
eq.~(\ref{studef}), {\it geometrical} moduli their real parts, {\it axions} their imaginary parts. 
When the remaining scalar excitations, living for example on branes, at brane 
intersections or at the orbifold fixed points, are consistently set to zero, the seven 
main moduli in eq.~(\ref{studef}) can be described by the K\"ahler potential
\beq
K= - \log \left( 
s \, t_1 \, t_2 \, t_3 \, u_1 \, u_2 \, u_3 
\right) \, .
\label{kahler}
\eeq
However, the identification of the real and imaginary parts in eq.~(\ref{studef}), in terms
of the spin-0 fluctuations of the ten-dimensional bosonic fields, is model-dependent. 

So far, the greatest effort went into identifying the effective superpotential $W$ for the main 
moduli as a function of the fluxes, and the consistency conditions on the latter: since $K$ is 
given by eq.~(\ref{kahler}), the knowledge of $W$ completely determines the F-term 
contribution to the scalar potential for the main moduli. A general result, in the approximation 
of negligible warping and in the field basis of eqs.~(\ref{studef}) and (\ref{kahler}), is that $W$ 
is a polynomial, at most of degree one in each of the seven main moduli, with specified relative
phases between the different monomials: this property may be ascribed to the underlying
$\Ne4$ supergravity structure \cite{dkpz}. 

In type-IIB orientifolds \cite{IIB} with O3/O7-planes, the only available fluxes surviving the 
orbifold and orientifold projections are those of the NSNS and RR 3-form field strengths. 
As a result, dilaton and complex-structure moduli, which in the standard notation correspond 
to $(S,U_1,U_2,U_3)$ in eq.~(\ref{studef}), can be stabilized on a flat background with 
spontaneously broken supersymmetry, but the K\"ahler moduli, corresponding to $(T_1,T_2,
T_3)$ in eq.~(\ref{studef}), remain as complex classical flat directions,  as in no-scale models  
\cite{noscale}. Similarly, type-IIB orientifolds with O5/O9-planes give, under suitable field
redefinitions, the same effective theory for the main moduli as heterotic orbifolds, thus 
full stabilization of the main moduli is impossible. To go further, perturbative or 
non-perturbative quantum corrections must be advocated. 

In the type-IIA theory,  a richer spectrum of possibilities emerges, thanks to the richer structure 
of NSNS, RR and geometrical fluxes  surviving the orbifold and orientifold projections 
\cite{dkpz,bc,kkp,lt,gl,vzIIA,dgkt,cfi}. As in the type-IIB case, we can obtain Minkowski vacua 
with unbroken or spontaneously broken $\Ne1$ supersymmetry, but there are always some 
residual classical flat directions for the geometrical moduli and their axionic superpartners. 
In contrast with the type-IIB case, however, we can also obtain $AdS_4$ vacua with all the 
geometrical moduli stabilized, as  explicitly shown in \cite{vzIIA} for the orbifold  $T^6 /( Z_2
\times Z_2)$ and later discussed for other compactifications in \cite{dgkt, cfi}. 

All the above results were derived ignoring the effects of the gauge fields localized on
D-branes, and the associated D-term contributions to the scalar potential, under the 
assumption that D-branes would preserve $\Ne1$ supersymmetry\footnote{In type-IIB 
constructions, we can generate supersymmetry-breaking configurations by considering 
brane-localized magnetic fluxes \cite{bachas}. Making use of T-duality, this is equivalent to 
type-IIA constructions with D6-branes at angles \cite{bdl}. $U(1)$ D~terms for 
$\Ne1$ type-II compactifications and their dependence on the geometrical moduli were 
previously discussed in \cite{DIIA,DIIAbis,bbkl,blt,cascur,jl}, and the possible r\^ole of magnetic 
fluxes in the stabilization of the bulk moduli in \cite{blt,cascur,magnetic}.}. As will be clear 
in the following, this is certainly consistent for the identification of supersymmetry-preserving 
vacua. However, a more careful investigation of brane-localized gauge fields and the
associated D~terms is needed to answer a number of questions.  We expect additional 
constraints on fluxes, originating from the generalized Bianchi identities (BI) for the gauge 
fields localized on D-branes. We also expect that D~terms contribute to the masses of 
geometrical moduli. If so, they could in principle remove the residual flat directions for 
the geometrical moduli that are typical of  Minkowski vacua (as suggested for example 
in \cite{gdm}), and perhaps decouple the masses of the geometrical moduli from the 
scale of the cosmological constant in the case of stable $AdS_4$ vacua. 
Also, we would like to check explicitly whether D~terms are allowed to relax to zero 
for the field configurations corresponding to supersymmetry-breaking vacua of the F-term 
potential. Finally, we would like to know whether D~terms can play a r\^ole in the generation 
of metastable de-Sitter vacua: a possible mechanism in the latter directions was recently 
proposed in \cite{vzprl} within a simple toy model,  but no explicit string realization of it 
exists so far. 

The main goal of the present paper is to examine the issues described
above (with the exception of the last one, whose clarification may require 
a more general theoretical framework). We stress that, as will be explained
in more detail later, $U(1)$ D~terms always give a non-vanishing contribution 
to the bulk moduli spectrum, in all compactifications where chiral matter fields 
arise from brane intersections (or from magnetized branes  in the mirror description).
Exploring the structure of D~terms is thus a mandatory step for the study of 
semi-realistic flux compactifications.

We now anticipate the main results of the paper. We explicitly derive, via
dimensional reduction of the D6-brane action, the expression for D terms
in O6 orientifolds of the type-IIA theory, compactified on the
$T^6/(Z_2\times Z_2)$ orbifold (consistently neglecting matter fields). 
We show that it agrees with the expected supergravity formula as 
long as a specific bound on the D~term and the gauge coupling is satisfied. 
We extend the results to other
$\Ne1$ string compactifications, of the type-IIB theory with O3/O7 or
O9/O5 orientifolds, and of the heterotic theory. We show that D terms are
compatible with fluxes only when the BI for the localized gauge fields are
satisfied, and we find the general form for these constraints. We extend
the discussion to the case of non-geometrical flux compactifications,
finding the corresponding modifications to the localized BI and to the
effective superpotentials. We show that gauge anomalies cancel
when both bulk and localized BI are satisfied. We clarify the r\^ole of the $U(1)$
D~terms in the problem of bulk moduli stabilization, both in $AdS_4$ and in
Minkowski vacua.

The plan of the paper is the following. We complete this introduction 
by recalling some basic facts about F and D~terms in $\Ne1$, $D=4$ 
supergravity, and by summarizing the results of \cite{vzIIA} for the effective
F-term potential and superpotential for the main moduli, in a simple but very interesting 
class of $\Ne1$ type-IIA compactifications.  We then discuss in section~2 the structure 
of the effective potential for the main moduli originating from a stack of $N$ D6-branes. We 
identify, in the appropriate limit, the F-term and D-term contributions, and the consistency 
conditions that must hold for the effective theory to be a standard $\Ne1$, $D=4$ supergravity. 
Since we restrict our attention to the functional dependence on the main moduli, setting all
other scalar fields to zero, the relevant D~terms are those associated with the $U(1)$ gauge
symmetries in the decompositions $U(N) \to U(1) \times SU(N)$, which act as shifts on the
four RR axions. In section~3 we discuss the localized BI for the gauge fields living on the 
D6-branes, the constraints they put on gauged symmetries and fluxes, and their intriguing 
relation with the Freed-Witten anomaly \cite{fw}. We also derive some consequences of the 
localized BI that may be relevant in other contexts, such as the discussion of non-geometrical 
fluxes \cite{dkpz, stw} and of gauge anomaly cancellation. In section~4 we comment on the 
extensions of our results to other $\Ne1$ compactifications, not only of the type-IIA theory
but also of the type-IIB and heterotic theories. We then discuss in section~5 the r\^ole of $U(1)$ 
D~terms in the problem of bulk moduli stabilization. We explain how D~terms can generate positive
contributions to the squared masses of some geometrical moduli that, albeit stabilized
on a supersymmetric $AdS_4$ vacuum, have negative squared masses from the F-term
potential. Finally, we describe some difficulties in moduli stabilization that arise on Minkowski 
flux vacua, with exact or spontaneously broken supersymmetry. We conclude in section~6
with a brief summary of our results and some comments on the prospects for future work.
Finally, in the Appendix we display some of our formulae for flux compactifications of the
type-IIA theory on the $T^6/(Z_2 \times Z_2)$ orbifold and O6 orientifold, in a more
explicit form that can be useful for model building.
 
%%%%%%%%%%%%%%%%%%%%%%%%%%%%%%%%%%%%%%%
\subsection{F and D~terms in $\Ne1$, $D=4$ supergravity}
We recall here some basic facts about the F- and D-term contributions to the scalar potential,
in a generic $\Ne1$, $D=4$ supergravity with  chiral multiplets $\phi^i \sim (z^i, \psi^i)$ and
vector multiplets $V^a \sim (\lambda^a, A_\mu^a)$.
Up to two derivatives in the bosonic fields, the gauge--invariant  supergravity
action is completely determined by three ingredients (see, e.g., \cite{Dbooks}). 
The first is the real and gauge-invariant K\"ahler function $G$, which can be written in terms 
of a real K\"ahler potential $K$ and a holomorphic superpotential $W$ as
\be 
G = K + \log |W|^2 \, .  
\label{ggen}
\ee
The second is the holomorphic gauge kinetic function $f_{ab}$, which
transforms as a symmetric product of adjoint representations, plus a
possible imaginary shift associated with anomaly cancellation. Generalized 
Chern-Simons terms may also be needed \cite{afl}, but they will not play any 
r\^ole in the situations discussed in this paper and will be neglected. Then
the generalized kinetic terms for the vector fields can be written as
\be
\label{gkin}
-\frac14 \, \tilde e_4 \, \Re f_{ab} \ F^a F^b + \frac12 \Im f_{ab} \ F^a \wedge F^b \, .
\ee
The third ingredient are the holomorphic Killing vectors $X_a = X_a^i (z) (\de/\de z^i)$, 
which generate the analytic isometries of the K\"ahler manifold for the scalar fields that 
are gauged by the vector fields. In the following it will suffice to think of $G$, $f_{ab}$ 
and $X_a$ as functions of the complex scalars $z^i$ rather than the superfields $\phi^i$.
The gauge transformation laws and covariant derivatives for the scalars in the chiral 
multiplets read
\be
\delta z^i  =  X_a^i \, \epsilon^a \, , 
\qquad
D_\mu z^i = \de_\mu z^i - A^a_\mu X_a^i \, ,
\ee
where $\epsilon^a$ are real parameters. The scalar potential is
\be 
\label{vgen}
V = V_F + V_D = e^G \l (G^i G_i - 3 \r) + \frac12 D_a D^a \, ,
\ee
where $G_i = \de G / \de z^i$, scalar field indices are raised with
the inverse K\"ahler metric $G^{i \ov{k}}$, gauge indices are raised
with $[(Re f)^{-1}]^{ab}$, and $D_a$ are the Killing potentials, real
solutions of the complex Killing equations:
\be 
X_a^i = - i \, G^{i \ov{k}} \, 
\frac{\de D_a}{\de \ov{z}^{\ov{k}}} \, .
\ee
The general solution to the Killing equation for $D_a$, compatible
with gauge invariance, is
\be
\label{eq:solD} 
D_a = i \, G_i \, X_a^i = i \, K_i \, X_a^i + 
i \, \frac{W_i}{W} \, X_a^i \, .  
\ee
It is not restrictive to assume that $K$ is gauge-invariant. If $W$ is also 
gauge-invariant\footnote{For $U(1)$ factors, $W$ can be gauge-invariant 
up to a phase. This corresponds to gauging the R symmetry and leads
to constant Fayet-Iliopoulos terms in $D_a$, whose possible 
phenomenological relevance was discussed in \cite{vzprl}. Since they 
play no r\^ole for the present paper, we will neglect here this possibility.},
\be
W_i \, X^i_a = 0 \, ,
\label{ginvw}
\ee
eq.~(\ref{eq:solD}) reduces to
\be 
\label{dred}
D_a = i \, K_i \, X_a^i \, .
\ee
For a linearly realized gauge symmetry, $i \, K_i \, X_a^i = - K_i \, (T_a)^i_{\; k} 
z^k$, and we recover the familiar expression of \cite{cremmer} for the D~terms. 
For a $U(1)$  axionic shift symmetry,
\be
X_a^i = i \, q_a^i \, ,
\label{fdfi}
\ee
where $q_a^i$ is a real constant. We thus obtain what are usually called, with a
slight abuse of language, field-dependent Fayet-Iliopoulos (FI) terms: they will 
play a crucial r\^ole in the rest of this paper.

We stress here two known consequences of  eq.~(\ref{eq:solD}), which shows 
that D~terms are actually proportional to F terms, $F_i = e^{G/2} \, G_i$. First, and in 
contrast with the rigid case, there cannot be pure D breaking of supergravity, unless 
the gravitino mass vanishes and the D-term contribution to the vacuum energy is
uncanceled, as in the (unrealistic) limit of global supersymmetry. Second, if 
$V_F$ admits a supersymmetric $AdS_4$ vacuum configuration, $\la G_i \ra = 0$
($\forall i$) and $\la e^G \ra \neq 0$, such configuration automatically minimizes 
$V_D$ at zero, and D~terms cannot be used to raise the vacuum energy from 
negative to positive or zero. Moreover, the 
gauge invariance of $W$, eq.~(\ref{ginvw}), puts severe constraints on the 
simultaneous presence of flux-induced superpotentials and D~terms: this
point will be discussed in detail in section~\ref{sec:lBI}. 

%%%%%%%%%%%%%%%%%%%%%%%%%%%%%%%%%%%%%%%
\subsection{F-term potential and superpotential for the main moduli}

In the following, we will mostly concentrate on type-IIA compactifications on the chosen
orbifold $T^6/(Z_2 \times Z_2)$, with the orientifold projection $\Omega (-1)^{F_L} I_3$, 
where $\Omega$ is the world-sheet parity operator, $(-1)^{F_L}$ is the space-time 
fermion number for left-movers, and $I_3$ acts as a parity on three of the six internal 
coordinates. For the main moduli of these compactifications, we know by now the 
effective superpotential and the full F-term contributions to the effective potential
\cite{dkpz, vzIIA},  as well as the consistency conditions on the fluxes coming from the BI
of the local symmetries gauged by bulk fields \cite{vzIIA}. To set the stage for our discussion 
of D~terms and localized BI, we briefly recall the main results of \cite{vzIIA}.

We begin with the bosonic field content of $D = 10$ type--IIA supergravity.  
In the NSNS sector,  we have the (string-frame) metric $g_{MN}$, the 2-form potential $B$ and 
the dilaton $\Phi$.  In the RR sector, we have the $(2k+1)$-form potentials $C^{(2k+1)}$, ($k=0,
\dots,4$), whose degrees of freedom are not all independent, being related by Poincar\'e duality. 

For definiteness, we  take the action of the orbifold projection on the internal coordinates $x^r$ 
($r = 5, \dots,10$) of the factorized 6-torus $T^6 = T^2 \times T^2 \times T^2$ to be
\beq
Z_2 \ : \;\; (z^1,z^2,z^3) \to (-z^1,-z^2,z^3) \, , 
\qquad
Z'_2 \ : \;\; (z^1,z^2,z^3) \to (z^1,-z^2,-z^3) \, , 
\eeq
where
\beq
z^1 = x^5 + i \ x^6 \, , 
\qquad 
z^2 = x^7 + i \ x^8 \, , 
\qquad 
z^3 = x^9 + i \  x^{10} \, ,
\eeq
and the action of the $I_3$ orientifold involution to be
\beq
I_3 \ : \;\; (z^1,z^2,z^3) \to  
(- \ov z^1, - \ov z^2, - \ov z^3) \, .
\eeq
The intrinsic parities of the bosonic fields with respect to the orientifold 
projection are: $+1$ for  $g_{MN}$, $\Phi$ and the RR $p$-form potentials with $p=3,7$; 
$-1$ for $B$ and the RR $p$-form potentials with $p=1,5,9$.

Before proceeding, we introduce some notation that will be used in the rest of the paper.
For the chosen toroidal orbifold, there are eight independent 3-cycles on the factorized
3-torus, four even and four odd under the orientifold projection.
Since each 3-cycle $\pi$ can be identified by a set of topological wrapping numbers,
\be
\pi = (m_1,n_1) \otimes (m_2,n_2) \otimes (m_3,n_3) \, ,
\label{pidef}
\ee
we can define the following basis for even and odd 3-cycles:
\bea 
\l \{ \barr{ccc}
\alpha^0&=&(1,0) \otimes (1,0) \otimes (1,0)\\
\alpha^1&=&(1,0) \otimes (0,1) \otimes (0,1)\\
\alpha^2&=&(0,1) \otimes (1,0) \otimes (0,1)\\
\alpha^3&=&(0,1) \otimes (0,1) \otimes (1,0)\earr\r. \, ,
\qquad
\l\{\barr{ccc}
\beta_0&=&(0,1) \otimes (0,1) \otimes (0,1)\\
\beta_1&=&(0,1) \otimes (1,0) \otimes (1,0)\\
\beta_2&=&(1,0) \otimes (0,1) \otimes (1,0)\\
\beta_3&=&(1,0) \otimes (1,0) \otimes (0,1)\earr\r. \, .
\label{basis}
\eea
Equivalently, we can define the dual basis $([\alpha]^I,[\beta]_I)$ ($I=0,1,2,3$)
for the cohomological 3-forms associated with the (even, odd) 3-cycles, 
normalized as
\beq
\int [\alpha]^I \wedge [\beta]_J=-\delta^I_J \, .
\label{norm}
\eeq
In the basis of eq.~(\ref{basis}), $\pi$ of eq.~(\ref{pidef}) and its dual 3-form
read
\be \label{pidefbis} 
\pi = p_I \, \alpha^I + q^I \, \beta_I \, , \qquad
[\pi]= p_I \, [\alpha]^I - q^I \, [\beta]_I \, ,
\ee
where
\bea \label{eq:defvec}
\l \{ \barr{ccc}
p_0&=&m_1 m_2 m_3\\
p_1&=&m_1 n_2 n_3\\
p_2&=&n_1 m_2 n_3\\
p_3&=&n_1 n_2 m_3\earr\r.  \, ,
\qquad
\l\{\barr{ccc}
q^0&=&n_1 n_2 n_3\\
q^1&=&n_1 m_2 m_3\\
q^2&=&m_1 n_2 m_3\\
q^3&=&m_1 m_2 n_3\earr\r. \, .
\eea
The action of $I_3$ is compatible with 
the presence of O6-planes with internal coordinates $(6,8,10)$, wrapping the 3-cycle 
$\pi_{06} = \alpha^0$. The action of the orbifold group induces additional  O6-planes in 
the three directions [$(6,7,9)$, $(5,8,9)$, $(5,7,10)$],  corresponding to 
$\pi_{06}^\prime = -\alpha^A$ $(A=1,2,3)$, respectively. 
A stack of parallel D6-branes wrapping a generic (factorizable) 
3-cycle $\pi$ will be characterized by eq.~(\ref{pidef}) or eq.~(\ref{pidefbis}). Later in 
this paper, we will also consider their mirror D6-branes with respect to the $O6$-plane 
$\pi_{O6}$, wrapping the 3-cycle:
\be 
\pi' =    p_I \, \alpha^I - q^I \, \beta_I   \,.
\ee

The imaginary parts of the seven main moduli are associated with the scalar 
components of the NS-NS  2-form potential and of the R-R 3-form potential 
surviving the chosen orbifold and orientifold projections. In the notation of \cite{vzIIA}:
\be
B_{56|78|910} = \tau_{1|2|3} \, , 
\qquad 
C^{(3)}_{6810} = \sigma \, , 
\qquad 
C^{(3)}_{679|589|5710} = - \nu_{1|2|3}  \, .
\label{axdef}
\ee
The real parts of the seven main moduli are associated with the invariant
components of the dilaton and the metric, traditionally decomposed as
\be
e^{-2\Phi} =  \frac{\hat s}{t_1 t_2 t_3} \, ,
\quad
 g_{MN}= {\rm blockdiag}\l ( \hat s^{-1}\, \widetilde{g}_{\mu\nu},
\ t_1 \,\hat u_1,\ \frac{t_1}{\hat u_1}, \ t_2 \, \hat u_2,\ 
\frac{t_2}{\hat u_2},\ t_3 \, \hat u_3,\ \frac{t_3}{\hat u_3} 
\r) \,, 
\label{metdec}
\ee
where $\widetilde{g}_{\mu\nu}$ is the metric in the $\De4$ Einstein
frame. However, the correct complexification of eq.~(\ref{studef}) is 
achieved only after the following field redefinition:
\be
s  =  \sqrt{\frac{\hat s}{\hat u_1\,\hat u_2\,\hat u_3}} \,, 
\quad 
u_1 = \sqrt{\frac{\hat s \ \hat u_2\,\hat u_3}{\hat u_1}} \, , 
\quad 
u_2 = \sqrt{\frac{\hat s \ \hat u_1\,\hat u_3}{\hat u_2}} \, , 
\quad 
u_3 = \sqrt{\frac{\hat s \ \hat u_1\,\hat u_2}{\hat u_3}} \, .
\label{redef}
\ee
Such redefinition, which can be inferred by computing the kinetic terms for the 
main moduli via dimensional reduction, is crucial for writing the effective $\Ne1$
supergravity in the standard form, and will play an important r\^ole also in the 
discussion of the D~terms\footnote{Analogous redefinitions, involving the $(\hat{s},
t_1,t_2,t_3)$ fields and leaving the $(\hat{u}_1,\hat{u}_2,\hat{u}_3)$ fields 
untouched, must be also performed in the case of type-IIB orientifolds.}.
Summarizing, the metric $\widetilde{g}_{\mu \nu}$ and the seven main moduli 
are the only bosonic bulk fields surviving the orbifold and orientifold projections.
There are no surviving zero modes for the bulk vector fields coming from the
ten-dimensional metric and from the $p$-form potentials in the NSNS and RR sectors.

We now list the various (constant) fluxes allowed by the chosen orbifold and 
orientifold projections. For the Scherk-Schwarz \cite{ss} geometrical fluxes 
$\omega$, we have twelve independent components: 
\be
(\omega_{68}^{\quad 10} \, ,  \, \omega_{10 6}^{\quad 8} \, , \, \omega_{8 10}^{\quad 6}) \, ; 
\quad
(\omega_{57}^{\quad 10} \, ,   \, \omega_{95}^{\quad 8} \, ,  \, \omega_{79}^{\quad 6}) \, ; 
\quad 
(\omega_{58}^{\quad 9} \, ,  \, \omega_{89}^{\quad 5} \, ,  \, \omega_{67}^{\quad 9} \, , 
\, \omega_{96}^{\quad 7} \, ,  \, \omega_{105}^{\quad 7} \, , \, \omega_{710}^{\quad 5})
\, .
\ee
In the following, we will adopt the conventions of \cite{vzIIA} for the contraction of
$\omega$ with any $p$-form. For the NSNS 3-form field strength $\ov{H}$, we 
have four independent components:
\be
\ov{H}_{579} \, ;
\quad
(\ov{H}_{5810} \, ,  \, \ov{H}_{6710} \, ,  \, \ov{H}_{689})
\, .
\ee
They can also be decomposed in the basis of eq.~(\ref{basis}). Finally, for the field strengths 
in the RR sector we have eight independent components: 
\be
\ov{G}^{(0)} \, ; \quad (\ov{G}^{(2)}_{56} \, , \,  
\ov{G}^{(2)}_{78} \, ,  \, \ov{G}^{(2)}_{910}) \, ;
\quad  (\ov{G}^{(4)}_{5678} \, ,  \, \ov{G}^{(4)}_{78910} \, ,  \, \ov{G}^{(4)}_{56910}) \, ;
\quad  \ov{G}^{(6)}_{5678910} \, .
\ee
The above fluxes must satisfy generalized BI associated with the local symmetries gauged 
by bulk fields, taking into account the possible existence of localized sources such as 
D6-branes and O6-planes. The integrability conditions of such BI provide the following 
non-trivial constraints on the allowed fluxes:
\beq
\omega \ \omega =0 \, , 
\label{eq:BIcH}
\eeq
\be
\frac12 \l ( \omega \, \ov G^{(2)}+\ov H \, \ov G^{(0)}\r)
= \sum_a N_a \, \mu_a \ [\pi_a] \, ,
\label{bicomp}
\ee
where in the last equation the index $a$ runs over the stacks of D6-branes and the O6-planes,
$N_a$ is the number of branes in each stack ($N_a=2^3$ for O6-planes), and $\mu_a$ is the RR 
charge\footnote{In our normalization ($\kappa_{10}^2=1$),
$T_a = \mu_a=1/2$ for the D6-branes and their images, which should be counted 
separately, and $T_a = \mu_a=-2$ for the O6-planes.}.

The effective superpotential is,  in compact geometrical form:
\be
W = \frac14 \int_{{\cal M}_6} \ov{\mathbf{G}} \, e^{i J^c}
- i \, \l( \ov H -i \omega \, J^c \r) \wedge \Omega^c \, ,
\label{eq:superpot}
\ee
where we have grouped the RR fluxes into the formal sum
$\mathbf{\ov G} = \sum_{p = even} \ov{G}^{(p)}$
and, in our conventions, $J^c$ and $\Omega^c$ read
\bea
J^c & = & J + i \ B \, ,
\qquad
J = {i \over 2} \sum_{A=1}^3 dz^A \wedge d \ov{z}^A \, ,
\nn \\
\Omega^c & = & \Re \left( i \ e^{\dd - \Phi} \ \Omega \right) 
+ i \ C^{(3)} \, , \qquad
\Omega = dz^1 \wedge dz^2 \wedge dz^3 \, .
\label{omegadef}
\eea
In particular, in our case:
\be
J^c_{56|78|910} = T_{1|2|3} \, , 
\qquad
\Omega^c_{6810} = S \, , 
\quad 
\Omega^c_{679|589|5710}= - U_{1|2|3} \, .
\label{jdef}
\ee

As discussed in \cite{vzIIA}, the contributions to the effective potential coming from
localized sources and the integrability conditions of eq.~(\ref{bicomp}) are crucial
for establishing the exact correspondence between the potential obtained via
generalized dimensional reduction and the standard $\Ne1$ formula for the
F-term potential,
\beq
\label{eq:potw}
V_F = e^{K}\l [ \sum_{i=1}^7 \l|W_i (\varphi) + 
K_i W(\varphi) \r|^2 - 3 |W(\varphi)|^2\r] \, ,
\qquad
\varphi^{1,\ldots,7} = (S,T_1,T_2,T_3,U_1,U_2,U_3) \, ,
\eeq
where $W_i \equiv \de W / \de \varphi^i$, evaluated for the superpotential $W$ 
of eq.~(\ref{eq:superpot}), and  $K_i \equiv \de K /  \de \varphi^i$.

%%%%%%%%%%%%%%%%%%%%%%%%%%%%%%%%%%%%%%%
\section{D~terms from D6-branes}

In this section we will derive the effective potential for the main moduli generated 
by a stack of $N$ D6-branes wrapping a generic factorizable 3-cycle
$\pi$, as in eq.~(\ref{pidef}), in the chosen class of type-IIA compactifications. 
The generalization to other $\Ne1$ string compactifications 
is discussed in section~\ref{sec:extensions}.
If the D6-branes are parallel to the O6-plane defined by $\pi_{O6}$, and we use the index 
$\alpha$ for the D6-brane internal space, the index $\hat \alpha$ for the orthogonal 
space, the embedding of the D6-brane world volume ($x^\alpha)$ into the 
ten-dimensional space ($X^M$) can be defined by
\be
X^M = x^\alpha \, \delta^M_\alpha +  \, \phi^{\hat \alpha} (x^\alpha) 
\, \delta_{\hat \alpha}^M \, ,
\ee
where $\alpha=\mu,6,8,10$, $\hat \alpha=5,7,9$, and the $\phi^{\hat \alpha}$ describe 
the brane fluctuations in the transverse directions. If instead the D6-branes wrap a generic 
3-cycle $\pi$, their embedding is described by:
\be
X^{\prime \, M} = \Lambda^M_{\quad N} \, X^N \, ,
\label{xpri}
\ee
where $\Lambda^M_{\quad N}$ is the rotation matrix (acting trivially on the four non-compact
dimensions):
\beq
\Lambda^M_{\quad N} =  \mathbf{{I}}_4 \, 
\bigotimes_{A=1}^3 \l ( \barr{c c} m_A & n_A \\ -n_A & m_A \earr \r)\, .
\eeq
We recall that the localized action for a stack of $N$ parallel D6-branes is $S_{DBI} + 
S_{WZ}$, where
\be
S_{DBI} = - N \, T_6 \, \int_{{\mathbb R}^4\times \pi} d^7 x \ e^{- \Phi} \, \sqrt{ - \det \l( g_{\alpha \beta} 
+ B_{\alpha \beta} + \, F_{\alpha \beta} \r) } 
\ee
is the Dirac-Born-Infeld (DBI) action and
\be
S_{WZ} = N \, \mu_6 \, \int_{{\mathbb R}^4\times \pi} \sum_{n=odd} A^{(n)} \ e^{F} \, ,
\qquad
\l[ A^{(n)} = e^B \, C^{(n)} \r] \, ,
\ee
is the Wess-Zumino (WZ) action. In the above equations  
we kept only the contribution from the $U(1)$ factor, in the decomposition
$U(N) \to U(1) \times SU(N)$ of the gauge group. 

To derive the D6-brane contributions to the effective potential for the seven main moduli,
it is sufficient to consider the tension term in the DBI action, keeping 
the dependence on the bulk metric and antisymmetric tensor but setting all the brane 
fluctuations to zero. Making use of eq.~(\ref{metdec}), we find:
\be
V_6 = N \,  T_6 \ \frac{e^\prime_7}{\tilde e_4} \,  e^{- \Phi} =
N \,  T_6 \  \frac{1}{{\hat s}^2} \sqrt{\hat s
\prod_{A=1}^3 \l(\frac{m_A^2}{\hat u_A}+n_A^2 \hat u_A\r)}  \, ,
\ee
where $e_7^\prime = \sqrt{- \det g_{\alpha \beta}^\prime}$ is the siebenbein determinant 
computed in the rotated coordinate system. Making use of the field redefinitions of eq.~(\ref{redef}), 
we observe that, in the conventions of eq.~(\ref{omegadef}), and after setting
\be
\widetilde{\Omega}_\pi = \int_\pi i \, e^{- \Phi} \, \Omega \, ,
\ee
we can write
\be
\Re \widetilde{\Omega}_\pi 
=  m_1 m_2 m_3 s - \sum_{A=1}^3  m_A n_B n_C u_A
= p_0 \, s - \sum_{A=1}^3  p_A \, u_A  \, ,
\label{reom}
\ee
\be
\Im \widetilde{\Omega}_\pi  =  \sqrt{s u_1 u_2 u_3} 
\l( \frac{n_1 n_2 n_3}{s} - \sum_{A=1}^3 \frac{n_A m_B m_C}{u_A} \r)
= \sqrt{s u_1 u_2 u_3} \l( \frac{q^0}{s} - \sum_{A=1}^3 \frac{q^A}{u_A} \r)  \, ,
\label{imom}
\ee
with $A\neq B\neq C=1,2,3$.
Then we obtain:
\be
V_6 =  \frac{N \, T_6}{s u_1 u_2 u_3} \sqrt{(\Re \widetilde{\Omega}_\pi )^2+(\Im
\widetilde{\Omega}_\pi )^2} \, .
\ee
To recover the standard form of the supergravity potential, it is useful to decompose $V_6$ as
\be
V_6 = V_{6F} + V_{D} \, ,
\ee
where
\bea
V_{6F} & = &  \frac{N \, T_6}{s u_1 u_2 u_3} \Re \widetilde \Omega_\pi \,, \\
V_D & = &  \frac{N \, T_6}{s u_1 u_2 u_3} \l ( \sqrt{(\Re \widetilde \Omega_\pi )^2+
(\Im \widetilde \Omega_\pi )^2}-\Re \widetilde \Omega_\pi  \r) \, .
\label{vdfirst}
\eea
This decomposition was already perfomed in \cite{bbkl}.
The contribution $V_{6F}$ was  a crucial ingredient, in the generalized dimensional 
reduction of \cite{vzIIA}, for reconstructing the full F-term contribution to the scalar potential, 
associated with the $\Ne1$ superpotential of eq.~(\ref{eq:superpot}) and the K\"ahler 
potential of eq.~(\ref{kahler}), after making use of the integrability conditions of 
eq.~(\ref{bicomp}). Notice in fact that, for supersymmetric D6-branes,
\beq \label{eq:susycond}
\Im \widetilde \Omega_\pi=0\,,\qquad \Re \widetilde \Omega_\pi > 0\,,
\eeq
$V_D=0$ and no other contribution
to the scalar potential, except $V_{6F}$, arises from the D6-brane action.

We will now show that $V_D$ can be identified, in an appropriate 
limit, with the $U(1)$ D-term part of  the supergravity potential.

We begin by observing that the universal $U(1)$ associated with the brane stack acts as a shift 
symmetry on the four RR axions $(\sigma,\nu_1,\nu_2,\nu_3)$ defined in eq.~(\ref{axdef}). To see 
this, it is sufficient to show that the kinetic terms for these four axions get suitably covariantized. 
The origin of this phenomenon is the WZ part of the D6-brane action, which contains the term
\beq 
\label{eq:AF}
N \, \mu_6 \  \int_{{\mathbb R}^4\times \pi}  A^{(5)} \wedge F \, ,
\eeq
where $F = dA$ is the 2-form field strength for the localized vector bosons. The term 
of eq.~(\ref{eq:AF}) is the only one linear in $A_\mu$, and with one derivative, that can 
contribute to the BI for $G^{(4)}$. The latter can be easily constructed from the dual 
formulation of \cite{vzIIA} and reads:
\beq
\frac{\delta}{\delta A^{(5)}} \l [ \int \frac12 A^{(5)} \wedge d G^{(4)} 
- N  \, \mu_6 \, \int_{{\mathbb R}^4\times \pi} A^{(5)} \wedge F \r] = 0 \, .
\eeq
Its solution gives the covariant derivatives for the main moduli, whose only non-trivial 
components are\footnote{Notice that, strictly speaking, there are two sets of vector 
bosons, the first associated with the stack of branes and the second associated with the 
mirror stack, and that the gauge field appearing in eq.~(\ref{covder}) is the antisymmetric 
combination of the corresponding $U(1)$ vectors, since the orthogonal combination is 
truncated away by the orientifold projection. }:
\be
\left\{
\begin{array}{lcccr}
G^{(4)}_{\mu6810} &  = &   \de_\mu \sigma - 2 \, N \, \mu_6 \,  n_1 \, n_2 \, n_3 \, A_\mu 
& = &   \de_\mu \sigma - 2 \, N \, \mu_6  \,  q^0 \,A_\mu  \\ 
G^{(4)}_{\mu679} &  = &  - \de_\mu \nu_1 - 2 \, N \, \mu_6  \,  n_1 \, m_2 \, m_3 \,A_\mu
& = & - \de_\mu \nu_1 - 2 \, N \, \mu_6  \,  q^1 \,A_\mu \\ 	
G^{(4)}_{\mu589} &  = & - \de_\mu \nu_2 - 2 \, N \, \mu_6  \,  m_1 \, n_2 \, m_3 \,A_\mu
& = & - \de_\mu \nu_2 - 2 \, N \, \mu_6  \,  q^2 \,A_\mu  \\ 	
G^{(4)}_{\mu 5710} &  = & - \de_\mu \nu_3 - 2 \, N \, \mu_6 \,  m_1 \, m_2 \, n_3 \,A_\mu 
& = & - \de_\mu \nu_3 - 2 \, N \, \mu_6  \,  q^3 \,A_\mu
\end{array}
\right. \, .
\label{covder}
\ee
Notice that the first line in eq.~(\ref{covder}) comes actually from a higher-derivative term, since
each $n_A$ ($A=1,2,3$) in the expression for $q^0$ can be seen as the flux for the 1-form 
$d \phi$. This is just the analog of the Green-Schwarz \cite{gs} higher-derivative term, $\int B 
\wedge F  \wedge F  \wedge F  \wedge F$, which gauges the shift symmetry of  the universal 
axion in the heterotic theory. Thanks to the gauging, flat axionic directions of the effective
potential involving the $\sigma$ and $\nu_A$ fields can be removed, since the corresponding
axions get absorbed into massive $U(1)$ vector fields, generalizing a known result for the
heterotic string \cite{dsw}. As noticed in \cite{cfi}, this mechanism 
allows to remove completely the residual axionic flat directions of the supersymmetric $AdS_4$ 
vacua, with all geometrical moduli stabilized, found in \cite{vzIIA,cfi}. From the covariant 
derivatives of eq.~(\ref{covder}) we can extract the corresponding  $U(1)$ Killing vectors:
\bea
i \, X^S &=& - 2 \, N \, \mu_6 \,  n_1\,n_2\,n_3 = - 2 \, N  \, \mu_6 \,  q^0 \,, \nn \\
i \, X^{U_A} &=& 2 \, N \, \mu_6 \,  n_A\,m_B\,m_C = 2 \, N \, \mu_6 \,  q^A \, ,
\label{killsu}
\eea
where $A \ne B \ne C = 1,2,3$.

We can also look at the terms of the effective action quadratic in the vector field
strengths, recalling that their standard form in $\Ne1$ supergravity is given by
eq.~(\ref{gkin}). Reducing the DBI action, we get:
\be
-  \frac14  \tilde e_4 \, N \, T_6 \, \sqrt{(\Re \widetilde \Omega_\pi)^2+(\Im
\widetilde \Omega_\pi )^2} \, F^2
= - \frac14 \,  \tilde e_4 \l( N \, T_6 \Re \widetilde \Omega_\pi 
+ s u_1 u_2 u_3\, V_D \r) \, F^2 
\label{fsquare}
\, .
\ee
Analogously, reducing the WZ action,  we get:
\be
\label{fftilde}
\frac12 N \, \mu_6 \  \l ( m_1 m_2 m_3 \, \sigma - 
\sum_{A=1}^3 m_A n_B n_C \, \nu_A \r) F \wedge F 
=
\frac12 N  \, \mu_6 \ \l( p_0 \, \sigma - \sum_{A=1}^3  p_A \, \nu_A \r) F \wedge F \, .
\ee

At first sight the effective theory, as derived above from dimensional reduction, does not seem 
to match the general structure of $\Ne1$ supergravity described in subsection~1.1. A disturbing 
fact is that the coefficient of $F^2$ in eq.~(\ref{fsquare}) cannot be seen as the real part of a 
holomorphic function $f$, as in eq.~(\ref{gkin}), whose imaginary part would be fixed by 
eq.~(\ref{fftilde}).  
Also, the $V_D$ contribution to the potential is not of the form dictated by eqs.~(\ref{vgen}), 
(\ref{eq:solD}) and (\ref{killsu}). The reason of this apparent discrepancy is the fact that the DBI 
action includes higher-derivative terms. In the case of generic angles, i.e. of D6-branes wrapping
a generic 3-cycle, the spontaneous breaking of $\Ne1$ supersymmetry cannot be described 
within the standard 2-derivative formulation of $\Ne1$ supergravity\footnote{For a discussion on 
how to include higher-derivative terms into the supergravity formalism, see \cite{bg}.}. Anyway,
the 2-derivative approximation was implicitly assumed when writing down the $D=10$
bulk effective action and reducing it to four dimensions. We must then look for a suitable 
limit in which higher-derivative terms coming from the DBI action can be neglected, and 
the D-brane contributions to the effective potential and to the kinetic terms for the gauge 
fields can be put in the standard $\Ne1$ supergravity form.
 
Exploiting the gauge-invariance of the superpotential\footnote{In section~\ref{sec:lBI}
we will show how this condition is guaranteed by localized Bianchi identities.},  
eq.~(\ref{ginvw}), we can directly compute the D~term through
eq.~(\ref{dred}), using the K\"ahler potential of eq.~(\ref{kahler}) and the Killing vectors of 
eq.~(\ref{killsu}):
\beq 
\label{eq:defD}
D =  N \, \mu_6 \  \l( \frac{n_1 n_2 n_3}{s}  - \sum_{A=1}^3  \frac{n_A m_B m_C}{u_A} \r)
=  N  \, \mu_6 \ \l( \frac{q^0}{s}  - \sum_{A=1}^3  \frac{q^A}{u_A} \r)  \, .
\eeq
Therefore eqs.~(\ref{imom}) and (\ref{eq:defD}) imply the identification
\beq
\Im \widetilde \Omega_\pi = \frac{\sqrt{s u_1 u_2 u_3} \ D}{N \, T_6} \, ,
\label{imombis}
\eeq
and allow us to rewrite eq.~(\ref{vdfirst}) as
\beq 
\label{eq:VD}
V_D= \frac12 \frac{1}{N \, T_6 \, \Re \widetilde \Omega_\pi}\,  D^2 \ \frac{2}{\sqrt{1+
\l(\frac{\Im \widetilde \Omega_\pi}{\Re \widetilde \Omega_\pi }\r)^2}+1} \, .
\eeq
For eq.~(\ref{eq:VD}) to be compatible with the standard formula of $\Ne1$ supergravity, 
eq.~(\ref{vgen}),  we must require that
\beq 
\label{eq:Dcond}
\l\vert \frac{\Im \widetilde \Omega_\pi}{\Re \widetilde \Omega_\pi}\r\vert \ll 1\, .
\eeq
In the limit of eq.~(\ref{eq:Dcond}), also the generalized kinetic terms for the $U(1)$ vector field,
eqs.~(\ref{fsquare}) and (\ref{fftilde}), assume their standard $\Ne1$ supergravity form, 
eq.~(\ref{gkin}), with:
\bea
f & = & N \, T_6 \ \l( m_1 m_2 m_3 S - m_1 n_2 n_3 U_1- n_1 m_2 n_3 U_2- n_1 n_2 m_3 U_3 \r) 
\nn \\ & = & N \, T_6 \ \l( p_0 \, S - \sum_{A=1}^3 p_A \, U_A \r) \, ,
\label{gkinfin}
\eea
in agreement with the results of \cite{DIIAbis} in the special case of supersymmetric D6-branes. 
Now, observing that 
\be
\Re \widetilde \Omega_\pi =
\frac{\Re f}{N \, T_6}  \, ,
\ee
and making use of eqs.~(\ref{reom}) and (\ref{imombis}), the limit in  
eq.~(\ref{eq:Dcond}) can also be rewritten as:
\beq \label{eq:Dbound}
\frac{\hat s}{\Re f}\, D =  g^2 \, D \, Vol_6 \; e^{-2\Phi} \ll 1\,,
\qquad
{\rm or}
\qquad
g^2 \, D\, M_P^2 \ll M_s^2 \, ,
\eeq
where $Vol_6=t_1\,t_2\,t_3$ is the volume of the internal manifold (in string units), 
$g^2=(\Re f)^{-1}$, and we have reintroduced explicitly the string mass scale 
$M_s$ and the four-dimensional Planck mass $M_P$ in the second expression.

Besides the condition in eq.~(\ref{eq:Dcond}) we implicitly assumed that
$\Re \widetilde \Omega_\pi>0$ throughout the derivation of the effective action.
This condition guarantees the softness of the supersymmetry breaking.
In fact, if $\Re \widetilde \Omega_\pi \le 0$ the tension term could not
be reabsorbed into the F-term potential nor be interpreted as a D~term.
We illustrate the various situations in figure~\ref{fig:Omega}.
Notice that the supersymmetric case of eq.~(\ref{eq:susycond}) is actually
disconnected (i.e. there is no soft limit) from the $\Re \widetilde \Omega_\pi < 0$
region, the origin being a singular point (vanishing D-brane volume).

\FIGURE[t]{
\epsfig{file=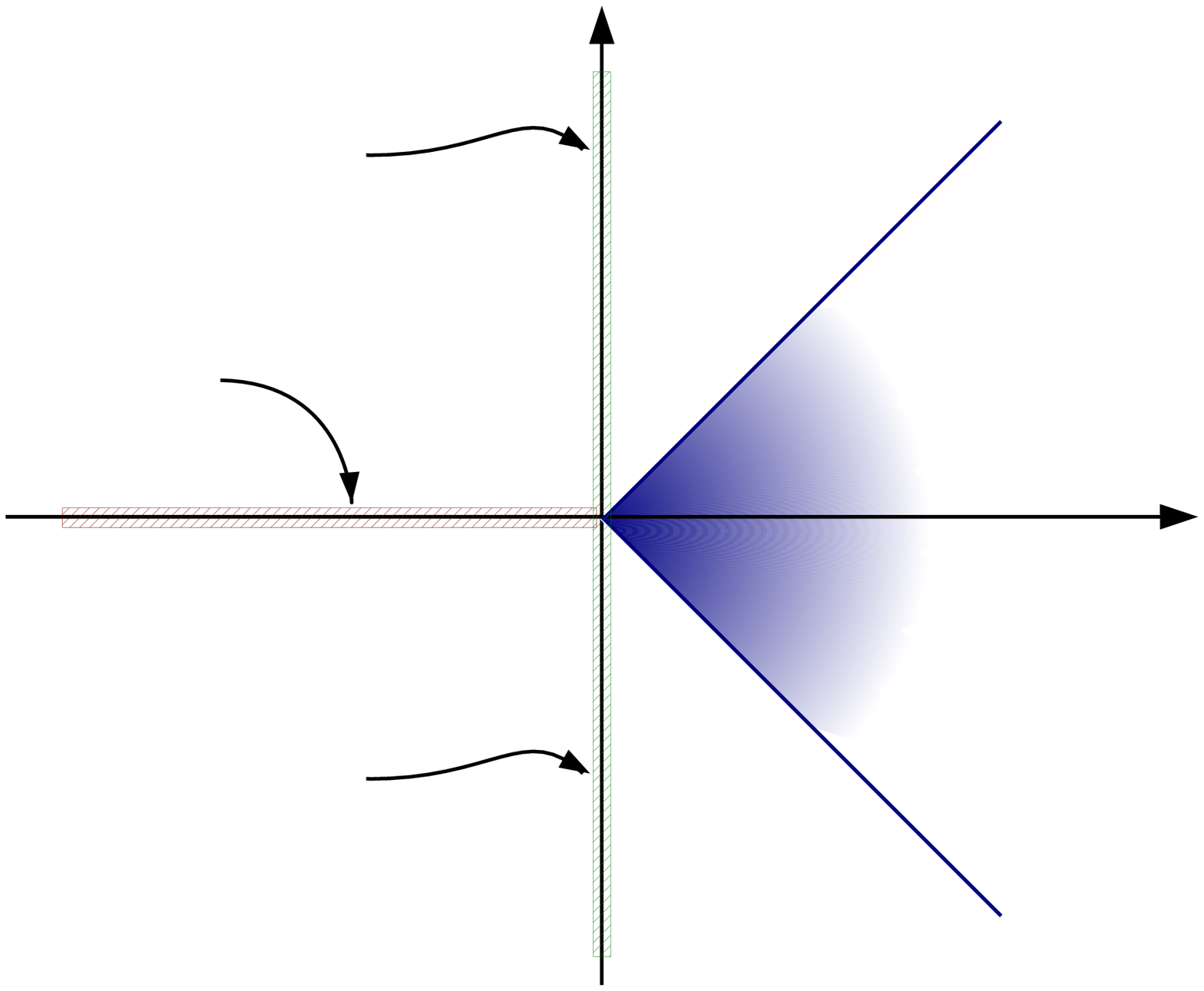,width=0.75\textwidth}
\vspace{10pt}
{\mbox{}\\[-160pt]\mbox{} \hspace{270pt}$\Re \widetilde \Omega_\pi$
\\[-134pt]\mbox{} \hspace{100pt} $\Im \widetilde \Omega_\pi$ 
\\[75pt]\mbox{} \hspace{200pt} $\Im \widetilde \Omega_\pi<\Re \widetilde \Omega_\pi$ 
\\[-20pt]\mbox{} \hspace{-130pt} \fbox{$\barr{c} \ov{D3}/\ov{D7} \\ (\ov{D9}/\ov{D5})\earr$}
\\[-95pt]\mbox{} \hspace{-74pt} \fbox{$\barr{c} {D9}/{D5} \\ ({D3}/{D7})\earr$}
\\[120pt]\mbox{} \hspace{-74pt} \fbox{$\barr{c} \ov{D9}/\ov{D5} \\ (\ov{D3}/\ov{D7})\earr$}
\\[30pt]}
\caption{D6-brane configurations in the complex $\widetilde \Omega_\pi$
plane. Supersymmetric configurations lie on the $\Re \widetilde
\Omega_\pi>0$ axis. The D6-brane DBI action still allows a low energy
supergravity description (see the text) in the shaded region where $\Im
\widetilde \Omega_\pi < \Re \widetilde \Omega_\pi$. Outside that region
supersymmetry is broken beyond the regime of validity of the effective
theory. For D6-branes with either $\Re \widetilde \Omega_\pi=0$ or $\Im
\widetilde \Omega_\pi=0$ and $\Re \widetilde \Omega_\pi<0$,  we indicate
the corresponding brane setup in the T-dual type-IIB O3/O7 (O9/O5)
orientifolds.
\label{fig:Omega}}
}

We conclude this section with some general comments. We recall that, given the 
generic factorizable 3-cycle $\pi$ of eq.~(\ref{pidef}), we can always associate 
to it an angle $\theta_A$ ($A=1,2,3$) in each factorized 2-torus:
\be
\tan \theta_A = \frac{n_A}{m_A} \hat u_A = \frac{n_A}{m_A} \sqrt{\frac{u_B u_C}{u_A\, s}} 
\, , \qquad (A \ne B \ne C =1,2,3) \, .
\ee
Brane configurations preserving $\Ne1$ supersymmetry are defined by \cite{bdl}:
\be
\theta_1 + \theta_2 + \theta_3 = 0 \, ,
\qquad
({\rm mod}~2 \pi) \, ,
\ee
or, equivalently:
\be
\sum_{A=1}^3 \tan \theta_A = \prod_{A=1}^3 \tan \theta_A \, .
\ee
Notice that the above conditions correspond to the vanishing of $\Im \widetilde \Omega_\pi$ 
and therefore of $V_D$, as expected for supersymmetry-preserving configurations. 

We can then distinguish among three different types of D6-brane configurations. Those
wrapping 3-cycles with $q^I=0$ ($I=0,1,2,3$) correspond to the four independent 
3-cycles $\alpha^I$ of the O6-planes and are always supersymmetric
as long as $\Re \widetilde \Omega_\pi>0$ (i.e. $p_0>0$, $p_A<0$): the
corresponding localized actions do contribute to the F-term potential but do not 
contribute to the D-term potential and, as we will see in the next section, are not
constrained by localized BI.  A more interesting class of D6-branes are those
with components $(p_I,q^I)$ along both even and odd 3-cycles. These D6-branes
do contribute to D~terms, can preserve $\Ne1$ supersymmetry only for specific values 
of the moduli,  can produce chiral matter fields
and can be used to absorb possible massless bulk axions via a
supersymmetric Higgs effect. As will be discussed in the following sections, they
can also play an important r\^ole in moduli stabilization via the associated D~terms,
but are strongly constrained by localized BI.
Finally, we have the D6-brane configurations with $p_I=0$ ($I=0,1,2,3$).  
They do not allow the condition of eq.~(\ref{eq:Dcond}) to be satisfied 
for any value of the moduli and correspond to having a supersymmetry-breaking scale lying 
beyond the range of validity of the effective supergravity theory.

%%%%%%%%%%%%%%%%%%%%%%%%%%%%%%%%%%%%%%%
\section{Localized Bianchi identities}
\label{sec:lBI}
In the previous section we showed how slightly decalibrated D6-branes can generate
D~terms depending on the dilaton and the complex structure moduli, associated 
to localized $U(1)$ vector multiplets that gauge some shift symmetry of the RR
axions:  the D-brane setup determines which axionic direction is gauged by the
$U(1)$. We also stressed that $U(1)$ gauge invariance imposes some non-trivial
conditions on D-brane configurations and fluxes: the effective superpotential $W$ 
must be invariant under the gauged axionic $U(1)$, eq.~(\ref{ginvw}). It was already 
noticed in \cite{cfi} that, in the absence of geometrical fluxes, eq.~(\ref{ginvw}) applied
to the superpotential of eq.~(\ref{eq:superpot}) can be understood in term of the 
cancellation of the Freed-Witten (FW) anomaly \cite{fw}:
\beq 
\label{eq:FW1}
\int_{\pi} \ov H=0\,,
\eeq
which can be interpreted in turn as the integrability condition for the BI of the localized 
field strength,
\beq \label{eq:lBI}
d F+H=0\,,
\eeq
evaluated on every 3-cycle $\pi$ wrapped by D6-branes. Since geometrical fluxes
can break some shift symmetries, we expect that in the presence of geometrical 
fluxes also eq.~(\ref{eq:FW1}) is modified: we will now derive the appropriate
modification, and relate it to the gauge invariance of the superpotential $W$. 

In the presence of several stacks of  D6-branes, the expression of 
eq.~(\ref{covder}) for the $U(1)$-covariant derivatives of the RR axions generalizes to:
\beq
D_\mu \sigma = \de_\mu \sigma - 2  \sum_a N_a \, \mu_a \, q^0_a \, A_\mu^a \, ,
\qquad
D_\mu \nu_A =  \de_\mu  \nu_A + 2 \sum_a N_a \, \mu_a \, q^A_a \, A_\mu^a \, ,
\quad
(A=1,2,3) \, ,
\label{eq:covder2}
\eeq
where the 3-cycles $\pi_a$, wrapping numbers ($m_A^a,n_A^a$) and four-vectors 
($p_{a \, I},q^I_a$) are associated with the $a$-th stack of $N_a$ D6-branes. Among all 
the $U(1)$ vectors appearing in eq.~(\ref{eq:covder2}), at most four combinations can get 
a mass via the St\"uckelberg mechanism. This number, however, can be reduced in the 
presence of fluxes.

First we consider the case without geometrical fluxes. The $U(1)$ shift symmetries that can be 
gauged are only those that leave the superpotential $W$ of eq.~(\ref{eq:superpot}) invariant. 
Since they act on the fields as follows
\beq
\delta \Omega^c = i\, \delta C^{(3)}=i\, \chi \,, 
\eeq
where $\chi=\chi^I [\beta]_I$ is a constant 3-form, this means
\beq 
\label{eq:invW}
\delta W=\frac14\int \ov H \wedge \chi \quad \propto \quad \ov H_I\,\chi^I=0\,,
\eeq
which is a condition on the Killing vectors $\chi^I$. Eq.~(\ref{eq:FW1}) also reads
\beq 
\label{eq:FW2}
\int \ov H \wedge [\pi_a]=\ov H_I\,q_a^I=0\,, 
\eeq
so that its solution $q_a^I$ is also solution of eq.~(\ref{eq:invW}), i.e. $q_a^I\propto \chi^I$. In 
other words, the only D6-branes allowed by the localized BI are those satisfying eq.~(\ref{eq:FW2}), 
which produce in eq.~(\ref{eq:covder2}) a gauging compatible with the symmetries of 
the superpotential, eq.~(\ref{eq:invW}). In the absence of geometrical fluxes $\omega$, only the 
flux $\ov H$ contributes to the FW anomaly cancellation condition, thus three axionic shift 
symmetries are always present and at least three independent $U(1)$ are needed if we want
to absorb the corresponding axions via a supersymmetric Higgs effect.

We now discuss the condition that guarantees gauge invariance of the effective theory in the
presence of geometrical fluxes. The constraint in eq.~(\ref{eq:invW}) now becomes:
\beq 
\label{eq:invWt}
\delta W=\frac14\int (\ov H-i \omega J^c)\wedge \chi=0\,, 
\eeq
and must be satisfied for every choice of the 2-form $J^c$ [see eq.~(\ref{jdef})].
Eq.~(\ref{eq:invWt}) gives now four constraints that in general are independent. The first 
one is just eq.~(\ref{eq:invW}), while the other ones, using the fact that $J^c\wedge \chi=0$,
can be rewritten as
\beq
\int J \wedge \omega \, \chi=0 \,, 
\eeq
for every value of $J$ in eq.~(\ref{omegadef}), or, equivalently, as
\beq 
\label{eq:invW2}
\int_\gamma \omega \, \chi=0\,,
\eeq
for every holomorphic 4-cycle $\gamma$ (5678, 78910, 56910). 

Summarizing, eq.~(\ref{eq:invWt}) gives four linear equations in the four variables 
$\chi^I$. If these equations are independent, there are no Killing vectors. In this case, 
all RR axions are stabilized by fluxes and no shift symmetry can be gauged. This also 
means that D6-branes can only wrap $q^I_a=0$ cycles, i.e. supersymmetric cycles, 
with identically vanishing D~terms. In this case, however, no chiral fermions 
arise from D-brane intersections. When two or more equations are dependent, 
then at least one axionic symmetry is left and D~terms 
are allowed. This is the case for the stable supersymmetric $AdS_4$ vacua found in 
\cite{vzIIA},  where all D~terms vanish on the vacuum because of supersymmetry.

By looking at the localized BI of eq.~(\ref{eq:lBI}), it is not clear how to derive the 
constraints on D6-branes needed to satisfy gauge invariance, eq.~(\ref{eq:invWt}). 
The authors of ref.~\cite{cfi} proposed to extend the FW anomaly cancellation 
conditions of eq.~(\ref{eq:FW1}) to
\beq 
\label{eq:fwcfi}
\int_\pi \ov H-i \omega J^c =0 \, ,
\eeq
evaluated on the vacuum. We now give a proof of this condition, using T-duality,
directly from the localized BI of eq.~(\ref{eq:lBI}). What we find, however, is a stronger 
condition: eq.~(\ref{eq:fwcfi}) must be satisfied for any $J^c$, to ensure gauge 
invariance as in eq.~(\ref{eq:invWt}).

We start by considering the localized BI in the type-IIB O9-orientifold with (magnetized)
D9-branes on twisted tori:
\beq 
dF+\omega F+H=0\,.
\eeq
The integrability condition reads:
\beq \label{eq:FWIIB0}
\omega \ov F+\ov H=0\,,
\eeq
evaluated on every invariant 3-cycle. Because of the orientifold $\ov H=0$, and 
without loss of generality we can rewrite this constraint as
\beq 
\label{eq:FWIIB}  
\int_\Sigma \omega \, \ov F = \int (\omega \ov F) \wedge [\Sigma] = 0  \,,
\eeq
for every 3-cycle $\Sigma$. We thus have eight independent constraints,
one for each independent 3-cycle ($\alpha^I$, $\beta_I$). We now perform
a T-dualization along the three directions of the torus that map the type-IIB 
O9-plane into the type-IIA O6-plane lying on the 3-cycle $\alpha^0$. Under 
this T-duality, magnetized D9-branes map into D6-branes at angles.  In
particular, the magnetic fluxes in each 2-torus T$^2_A$ will determine the 
D6-brane embedding via the relation
\beq 
\label{eq:Fnm}
\int_{T^2_A} \ov F = \frac{n_A}{m_A}\, .
\eeq
Therefore, T-duality maps $\ov F$ of eq.~(\ref{eq:FWIIB}) into pull-backs 
(or pull-forwards) of type-IIA. In the NSNS sector, T-duality mixes
geometrical fluxes $\omega$ with $\ov H$-fluxes and, in general, also with
non-geometrical fluxes (see \cite{dkpz, stw}). Out of the eight constraints
of eq.~(\ref{eq:FWIIB}), four (corresponding to $\Sigma=\beta_0$ and 
$\Sigma=\alpha^A$) give constraints on the non-geometrical fluxes of type-IIA,
and will be discussed below in subsection~\ref{sec:non-geo}. We discuss now the 
other four constraints. In the case $\Sigma=\alpha^0$, T-duality transforms the 
components of $\omega$ of type-IIB into $\ov H$ of type-IIA, and eq.~(\ref{eq:FWIIB}) 
into the constraint of eq.~(\ref{eq:FW1}):
\beq 
\label{eq:FW1b}
\int_{\pi} \ov H=0 \, ,
\eeq
where $\pi$ is a D6-brane with wrapping numbers determined by eq.~(\ref{eq:Fnm}). 
Notice that not all the components of eq.~(\ref{eq:FW1b}) can be recovered from 
eq.~(\ref{eq:FWIIB}), in fact the component $\ov H_0$ is lacking. The reason is that
this component is T-dual to a non-geometrical flux of type-IIB, which we did not turn on
in eq.~(\ref{eq:FWIIB}). In the case $\Sigma=\beta_A$, $\omega$ is mapped into 
itself by T-duality and eq.~(\ref{eq:FWIIB}) into
\beq 
\label{eq:FW3}
\int_\gamma \omega [\pi]=0\,,
\eeq
for each of the three invariant 4-cycles $\gamma$. Again, one component of $\omega$ 
is missing in each of the three conditions of eq.~(\ref{eq:FW3}), because it would be 
T-dual to a non-geometrical flux of type-IIB, which was kept turned off in eq.~(\ref{eq:FWIIB}). 
%Add%%%%%%%%%%%%%%%%%%%%%%%%%%%%%%%%%%%%%%%%%%
The condition in eq.~(\ref{eq:FW3}) has actually a geometrical interpretation\footnote{We thank
Alessandro Tomasiello for a discussion on this point.}. The geometrical fluxes $\omega$ change 
the topology of the internal manifold by removing some of the internal cycles. Eq.~(\ref{eq:FW3})
is just the condition that the 3-cycle $\pi$ wrapped by the D6-brane be a true cycle, in fact:
\beq
0=\int_\gamma \omega[\pi]=\int_\gamma d [\pi]=-\int_\pi d [\gamma]=-\int_{\partial \pi} [\gamma] 
\ \Rightarrow \ \partial \pi = \emptyset \, .
\eeq
%
%%%%%%%%%%%%%%%%%%%%%%%%%%%%%%%%%%%%%%%%%%%%%
The 1+3 constraints of eqs.~(\ref{eq:FW1b}) and (\ref{eq:FW3}) can thus be rewritten as follows:
\beq 
\label{eq:FW4}
\int_\pi (\ov H-i\,\omega J^c) = 0 \,, 
\eeq
which, repeating the discussion given in the absence of geometrical fluxes, is 
precisely what we need to require that only axionic directions leaving the 
superpotential invariant can be gauged. 

We can now understand how $U(1)$ D~terms for the $(s,u_1,u_2,u_3)$ moduli
are strongly constrained by localized BI. Gauge invariance of the flux-induced
superpotential implies
\beq
D_a =  - 2 \, N_a \, \mu_6 \, (q^0_a \, K_S - q^A_a  K_{U_A} ) \,,
\eeq
with strong constraints linking the fluxes and the $q^I$, coming from localized BI. 
For instance,
in the case of generic geometrical fluxes only one axionic shift symmetry is
preserved by $W$, thus the vector $q^I_a$ is determined up to a normalization 
factor, and the form of the D~term is completely specified by the bulk fluxes, without
need of making explicit reference to D-brane data!

%%%%%%%%%%%%%%%%%%%%%
\subsection{Consistency conditions for non-geometrical fluxes}
\label{sec:non-geo}

We showed above that,  by applying T-duality to the localized BI in the
type-IIB theory, we get, besides the constraints for $\ov H$ and $\omega$,
additional consistency relations connecting the D-brane setup with 
non-geometrical fluxes. The latter show up when T-duality is applied to 
the usual NSNS 3-form fluxes, as follows \cite{stw}:
\beq
\ov H_{mnr}	\, \stackrel {{\cal T}_r} \longleftrightarrow \,
\omega_{mn}^{\quad r}\, \stackrel {{\cal T}_n}  \longleftrightarrow \,
Q_{m}^{\quad  nr}\, \stackrel {{\cal T}_m} \longleftrightarrow \, 
R^{mnr}\,,
\eeq
where ${\cal T}_{r,m,n}$ indicates a T-dualization in the $(r,m,n)$ internal direction.

Consider now the cases $\Sigma=\beta_0$ and $\Sigma=\alpha^A$ in 
eq.~(\ref{eq:FWIIB}), which were neglected before. They transform $\omega$ 
into $R$ and $Q$ fluxes, respectively, and map eq.~(\ref{eq:FWIIB}) into
the 1+3 conditions
\bea 
\label{eq:FWng}
\Sigma=\beta_0 &:& \quad \qquad R\,[\pi]=0\,, \nn \\
\Sigma=\alpha^A&:& \quad \int [\gamma] \wedge Q\,[\pi]=0\,, 
\quad (\forall\,\gamma) \,,
\eea
where $R\,[\pi]$ and $Q\,[\pi]$ are a 0-form and a 2-form 
whose components read:
\beq
R\,[\pi]=R^{mnr}\,[\pi]_{mnr} \,, 
\qquad 
(Q\,[\pi])_{mn} = Q_{[m}^{\quad rs}\,[\pi]_{rsn]} \,.
\eeq
More generally, when applied to a $p$-form $X$, $R$ and $Q$ decrease its 
rank by 3 and 1, respectively. In our notation:
\bea
(R\,X)_{r_1\cdots r_{p-3}}&=&R^{s_1\,s_2\,s_3}X_{s_1\,s_2\,s_3\,r_1\cdots r_{p-3}}\,, \nn \\
(Q\,X)_{r_1\cdots r_{p-1}}&=&Q_{r_1}^{\quad s_1\, s_2} X_{s_1\,s_2\,r_2\cdots r_{p-1}}+ 
{\rm permut.} \label{eq:satng}
\eea
Notice also that, for consistency, the intrinsic orientifold parities for 
the NSNS fluxes must be chosen to be:
\beq
\ov H \ (-) \,, \qquad \omega \ (+)\,, \qquad Q \ (-)\,, \qquad R \ (+)\,.
\eeq

Eqs.~(\ref{eq:FW1b}), (\ref{eq:FW3}) and (\ref{eq:FWng}) can also be 
rewritten in a compact form as
\bea 
\label{eq:FW5}
&&\int e^{iJ} (\ov H+ \omega + Q + R)[\pi] \nn  \\
&= &\int \ov H \wedge [\pi]+ i\, J\wedge \omega [\pi]- 
\frac12 J\wedge J\wedge Q[\pi] - \frac{i}{3!}\, J\wedge J\wedge J\, R[\pi]=0\,,
\quad 
(\forall \, J) \,.
\eea

The constraints of eq.~(\ref{eq:FWng}) are consistency conditions for
the simultaneous presence of D-branes and non-geometrical fluxes, and
must be added to the constraints from the bulk BI found in \cite{stw} 
when these type of fluxes are considered.
Still they guarantee that the modified non-geometrical superpotential:
\beq \label{eq:superpotng}
W=\frac14\,\int e^{iJ^{c}} \l( {\bf \ov G}-i\,(\ov H+ \omega + Q + R)\Omega^{c} \r)
\eeq
be invariant under the axionic shift symmetries gauged by the D6-branes.

Analogously, we can start from eq.~(\ref{eq:FW5}) and T-dualize it back to 
type-IIB, to extend eq.~(\ref{eq:FWIIB}) to non-geometrical fluxes.
This case will be discussed further in section~\ref{sec:extensions}.

%%%%%%%%%%%%%%%%%%
\subsection{Generalized Bianchi identities and gauge anomaly cancellation}
\label{sec:anom}

Localized BI are important not only to guarantee the tree-level
gauge invariance of the superpotential but, together with the bulk BI,
they also play a crucial r\^ole in the cancellation of the gauge anomalies.
To see how anomaly cancellation works in the presence of 
fluxes\footnote{We thank Angel Uranga for discussions on this point.}, 
we can generalize the standard proof (see e.g. \cite{uranga,marchesano}) 
valid in the absence of fluxes.

In toroidal compactifications, 
the cubic $SU(N_a)^3$ anomaly $\cA_a$ is proportional to the number of 
chiral fermions charged with respect to $SU(N_a)$, i.e.
to the sum of all D-brane intersections with the D-brane stack $a$,  namely:
\beq 
\label{eq:anomSU3-tor}
\cA_a= \sum_b I_{ab} N_b\,,
\eeq
where the intersection number,
\beq
I_{ab}=\int [\pi_a] \wedge [\pi_b]=p_{a\, I}\, q^I_b- q_a^I\, p_{b\, I}\,,
\eeq
counts the number of intersections with the $b$-th stack of D-branes.
Notice that, in order to have chiral matter, i.e. $I_{ab}\neq0$, at least 
one stack of D-branes must have $q^I\neq 0$, thus a non-trivial D term.

In constructions involving orbifolds and orientifolds, eq.~(\ref{eq:anomSU3-tor}) 
is modified into:
\beq 
\label{eq:anomSU3}
\cA_a= \sum_b \mu_b I_{ab} N_b\,,
\eeq
where the sum runs over both D-branes and O-planes.
This is due to the contributions from non-fundamental 
representations induced by the O-planes (see e.g. \cite{marchesano}).
By using the bulk BI in the presence of fluxes, eq.~(\ref{bicomp}),
the anomaly reads
\beq
\cA_a= \int [\pi_a] \wedge \sum_b \mu_b N_b[\pi_b]=\frac12 \int [\pi_a] \wedge 
\l (\omega \ov G^{(2)}+\ov G^{(0)} \ov H \r)\,,
\eeq
or equivalently
\beq
\cA_a\propto \int_ {\pi_a} \l ( \omega \,\ov G^{(2)}+\ov G^{(0)} \ov H\r) \,,
\eeq
which vanishes because of the localized BI (\ref{eq:FW4}).

After having imposed localized and bulk BI as above, also
mixed $U(1)$-$SU(N_a)^2$ and $U(1)^3$ gauge anomalies 
cancel via the generalized Green-Schwarz mechanism \cite{marchesano,uranga,bcls},
arising from the gauging of the shift symmetries discussed before.

%%%%%%%%%%%%%%%%%%%%%%%%%%%%%%%%%%%%%%%
%%%%%%%%%%%%%%%%%%%%%%%%%%%%%%%%%%%%%%%

\section{Extensions to other $\Ne1$ compactifications}
\label{sec:extensions}

The results obtained in the previous sections were derived explicitly in 
a particular example, the $T^6/(Z_2\times Z_2)$ orbifold of the type-IIA 
O6 orientifold. However, they are also valid in more general compactifications.
This is due to the fact that the structure of D terms and BI is mainly 
determined by gauge invariance and supersymmetry.

An example of the previous statement is the superpotential of  eq.~(\ref{eq:superpot}). 
Derived explicitly in the $T^6/(Z_2\times Z_2)$ orbifold case \cite{vzIIA}, it seems to
hold for all known geometrical $\Ne1$ type-IIA compactifications \cite{dkpz,gl,vzIIA,dgkt,cfi}. 
The details of each specific compactification only affect the number of active
moduli inside $\Omega^{c}$ and $J^{c}$, and the allowed components for the 
fluxes $\bf \ov G$, $\ov H$ and $\omega$. Analogously, the general formulae for the 
bulk [eq.~(\ref{bicomp})] and localized [eqs.(\ref{eq:FW1b}) and (\ref{eq:FW3})] BI 
apply to every $\Ne1$ compactification of the type-IIA theory on the O6 orientifold.
The particular flux and O6-plane content are, however, model-dependent.
Finally, from supersymmetry and gauge invariance, we know that D~terms are 
determined  just by the Killing vectors and by the K\"ahler potential [eq.~(\ref{dred})]. 
In particular, the former are fixed by the RR-couplings of eq.~(\ref{eq:AF}). When the 
latter reduces to a form analogous to eq.~(\ref{kahler}), as in \cite{dkpz,gl,vzIIA,dgkt,cfi}, 
then the functional dependence of the D~terms on the closed string moduli is fixed
to be the one of eq.~(\ref{eq:defD}).

In this way, the connection between BI, gauge invariance of
the superpotential, anomaly cancellation and D terms easily
extends beyond the specific orbifold and orientifold explicitly considered in 
the previous sections. The same arguments can be used to show 
that the extensions of the localized BI [eq.(\ref{eq:FW5})] and of the 
superpotential [eq.(\ref{eq:superpotng})] to non-geometrical fluxes 
must hold in general.

The discussion can also be extended to type-IIB compactifications. 
D~terms have already been computed, 
in compactifications with magnetized branes, in a number 
of cases  \cite{blt,cascur,jl}. Since the derivation is analogous to the type-IIA case,
we will not repeat it here. We just give the results, focusing on the intriguing
connection between gauge invariance and localized BI.

In the type-IIB  case, the localized BI can be obtained by mirror symmetry
from our eq.~(\ref{eq:FW5}), and read:
\beq \label{eq:FWIIB0ng}
\ov H + \omega\,\ov F + \frac12 Q\, (\ov F\wedge \ov F) + 
\frac{1}{3!} R\, (\ov F\wedge \ov F\wedge \ov F)=0\,,
\eeq
evaluated on every 3-cycle wrapped by a D-brane. The indices of 
$Q$- and $R$-fluxes are saturated as described in eq.~(\ref{eq:satng}).
Without non-geometrical fluxes, eq.~(\ref{eq:FWIIB0ng}) 
reduces to eq.~(\ref{eq:FWIIB0}). Moreover, only part of the
terms of eq.~(\ref{eq:FWIIB0ng}) survive the orientifold projections.
In particular, in O9/O5 compactifications $\ov H$ and $Q$ vanish,
while in O3/O7 compactifications $\omega$ and $R$ vanish.

In the O9/O5 case, D9-branes fill the whole ten-dimensional space
and, as expected, the localized BI correspond to the bulk BI of the
type-I or heterotic theory \cite{km}:
\beq
\omega\,\ov F=0\,.
\eeq
This equation is precisely what is needed for the effective superpotentials
of the type-IIB O9 and heterotic \cite{hetsup,dkpz} theories,
\bea
W_{IIB(O9)}&\propto&\int \l ( \ov G^{(3)} - i \omega J^{c} \r )\wedge \Omega^{c}\,, \nn \\
W_{Het}&\propto&\int \l ( \ov H - i \omega J^{c} \r )\wedge \Omega^{c}\,,
\eea
to be gauge invariant under the shift symmetries
\beq
J^{c}\to J^{c} + X\, .
\eeq
In fact, the Killing vectors $X=i \ov F$ are associated to
the gauging of the shift symmetry produced by the coupling
\beq
\int A^{(6)}\wedge \ov F \wedge F\,,
\eeq
where $A^{(6)}$ is the RR 6-form $C^{(6)}$ dual to $C^{(2)}=\Im J^{c}$ 
in the type-IIB O9 theory, or the NSNS 6-form $B^{(6)}$ dual to 
$B^{(2)}=\Im J^{c}$ in the heterotic case.
In both cases, geometrical fluxes do not produce a potential
for the universal axion $\Im S=C^{(6)}\,(B^{(6)})$ of the type-IIB
(heterotic) theory, which is gauged by the Green-Schwarz term
\beq \label{eq:gsterm}
\int A^{(2)}\wedge \ov F \wedge \ov F \wedge \ov F \wedge F\,.
\eeq
On the other hand, the corresponding Killing vector $\ov F \wedge \ov F \wedge \ov F$
is constrained by the non-geometrical $R$-fluxes, which
in fact induce superpotentials for the chiral superfield containing the axion:
\bea
W_{IIB(O9)}&\propto&\int (\ov G^{(3)}-i\,\omega\,J^{c} -i\,R\,\tilde S) \wedge \Omega^{c} \, , 
\qquad \tilde S\doteq \star_{(6)}S  \,,\nn \\
W_{Het}&\propto&\int (\ov H-i\,\omega\,J^{c} -i\,R\,\tilde S) \wedge \Omega^{c} \, .
\eea
When interpreted in terms of an underlying gauged $\Ne4$ supergravity, 
these superpotentials exhibit  a non-trivial de~Roo-Wagemans phase \cite{dRW}.
In the general case, with both $\omega$ and $R$ fluxes turned on,
eq.~(\ref{eq:FWIIB0ng}) guarantees that the combination of axions
gauged by fluxes is associated to a flat direction of the superpotential.

As in the type-IIA case, D~terms can be derived taking the DBI action and
performing the same supergravity limit of eq.~(\ref{eq:Dbound}). 
The result is just the mirror symmetric of the type-IIA result 
of eq.~(\ref{eq:defD}), where the wrapping numbers
are replaced by magnetic fluxes via eq.~(\ref{eq:Fnm}) 
and the $U$ and $T$ moduli are exchanged, i.e.
\beq
D\propto\l(\frac{n_1\,n_2\,n_3}{s}-\sum_{A=1}^3 \frac{n_A m_B m_C}{t_A}\r)
=m_1 m_2 m_3 \,\l ( \frac{\prod_{A=1}^3 \ov F_A}{s}
-\sum_{A=1}^3  \frac{\ov F_A}{t_A} \r )\,.
\eeq
The linear terms in $\ov F$ can also be derived by generalized 
dimensional reduction, using the standard heterotic action. 
The cubic term, however, corresponds to higher-derivative contributions,
as the associated Green-Schwarz term of eq.~(\ref{eq:gsterm}).

In the case of type-IIB O3/O7 orientifolds, only $\ov H$ and $Q$ fluxes can be 
turned on. Without non-geometrical fluxes, the effective superpotential \cite{gvw},
\beq
\int ( \ov G^{(3)}-i \,S \,\ov H)\wedge \Omega^{c}\,,
\eeq
produces only a potential for the dilaton $S$ and the complex structure moduli $U$, 
while the K\"ahler moduli $T$ remain
undetermined. Accordingly, the localized BI of eq.~(\ref{eq:FWIIB0ng}) 
require the vanishing of the $\ov H$ flux over all possible 3-cycles 
wrapped by a D-brane. This condition is actually satisfied trivially by all 
branes but magnetized D9-branes, which in fact gauge the shift symmetry 
of the axion $\Im S=C^{(0)}$ through the coupling
\beq
\int C^{(8)}\wedge F\,.
\eeq
This means that magnetized D9-branes cannot coexist with $\ov H$-fluxes,
unless also non-geometrical fluxes are considered (see below).
As before, non-geometrical fluxes modify both the superpotential, which now reads
\beq
\int ( \ov G^{(3)}-i\, S\, \ov H - i Q \tilde J^{c})\wedge \Omega^{c}\,, 
\qquad \tilde J^{c} \doteq \star_{(6)} J^{c}\,,
\eeq
and the localized BI [see eq.~(\ref{eq:FWIIB0ng})]. This produces
a potential for the $T$ moduli in $J^{c}$, which  is however compatible 
with the gauging of the shift symmetries associated to $J^{c}$
and generated by magnetized D9- and D7-branes via the couplings
\bea
D9&:&\int C^{(4)}\wedge \ov F \wedge \ov F \wedge F\,,\nn\\
D7&:&\int C^{(4)}\wedge \ov F \wedge F\,.
\eea
Again, in the presence of both $\ov H$ and $Q$ fluxes the localized
BI guarantee that the axionic directions gauged by magnetized D-branes
are associated to shift symmetries of the effective superpotential.
Finally, D~terms have the form:
\beq
D\propto\l ( \frac{m_1 m_2 m_3}{s}-\sum_A \frac{m_A n_B n_C}{t_A}\r )\,,
\eeq
with $m$'s and $n$'s exchanged with respect to the former cases because of
T-duality. The first contribution comes from D9-branes [which must be magnetized
in order to satisfy eq.~(\ref{eq:Dbound})], while the others come
from magnetized D7- and D9-branes. 

Notice that also in the
type-IIB case, after the appropriate field redefinitions, the D~terms are
homogeneous functions of $(s,t_1,t_2,t_3)$. Mechanisms for moduli 
stabilization via D~terms that rely on a different functional dependence
on the geometrical moduli cannot be consistently implemented, at least
as long as matter fields are neglected. 

Summarizing, we can see that, independently of the choice of the ten-dimensional
theory and of the details of the compactification, there is a precise correspondence 
between gauge invariance of the effective flux superpotential and localized BI. This 
translates into strong consistency conditions for the coexistence of fluxes and D~terms.
Remarkably, the same connection holds also in compactifications with non-geometrical fluxes.

In analogy with the type-IIA case discussed in section~\ref{sec:anom},
also in the type-IIB case gauge chiral anomalies cancel 
automatically when both bulk and localized BI are satisfied.
The proof can be derived along the line of section~\ref{sec:anom}
and holds also when non-geometrical fluxes are turned on.

All these results can have important consequences for string model-building. 
For instance, in O3/O7 type-IIB compactifications on the $T^6/(Z_2\times Z_2)$ 
orbifold,  it is a standard lore to add magnetized D9-branes, to avoid the positivity 
bounds on the bulk BI coming from flux quantization. However, according to the
discussion above, these types of branes are incompatible with $\ov H$ fluxes in 
geometrical compactifications\footnote{This inconsistency was already noticed in 
\cite{cascur}, where it was proposed to overcome the problem by adding fractional 
D5-branes, acting as sources for the localized BI. This modification, however, cannot 
be introduced without additional modifications to the model, because of the connection 
between gauge invariance of the superpotential, anomaly cancellation, bulk and 
localized BI.}.

%%%%%%%%%%%%%%%%%%%%%%%%%%%%%
\section{D~terms and moduli stabilization}

In the previous sections we showed how D~terms arising from D-branes can 
be embedded, under some well-defined consistency conditions, into the 
standard formalism of $\Ne1$, $D=4$ supergravity. We now comment on the 
r\^ole of $U(1)$ D~terms, with their specific dependence on the geometrical 
moduli, for the problem of bulk moduli stabilization in string compactifications 
with fluxes. For definiteness we refer, as before, to the $T^6/(Z_2 \times Z_2)$ 
orbifold and the $(-1)^{F_L} \Omega I_3$ orientifold of the type-IIA theory, but 
the results have more general validity. For the sake of clarity, and for an 
easier use of our results for model building, we collect some of the formulae 
of sections~2 and~3, translated into a more explicit notation, in the Appendix.

A property of the stable $AdS_4$ supersymmetric vacua found in the flux 
compactifications of \cite{vzIIA, cfi} is the existence of modes with negative squared
masses. Their presence is not a problem, because they satisfy the Breitenlohner-Freedman
bound \cite{BF} that controls stability in $AdS$ spaces. These "$AdS$ tachyons" are 
usually given by the linear combinations of geometrical moduli "parallel" to the linear 
combinations of axions gauged by the $U(1)$ D6-brane vectors. As explained in 
section~2, this dependence has a precise relation with the functional dependence 
of the $U(1)$ D~terms on the geometrical moduli. In this case, therefore, there can be 
a twofold r\^ole played by the $U(1)$ gauge interactions associated with D6-branes and 
by the corresponding D~terms, for the problem of bulk moduli stabilization. On the one 
hand, the gauged axionic flat directions of the F-term potential are removed via the 
St\"uckelberg mechanism: a supersymmetric Higgs effect takes place and the axion 
provides the longitudinal degree of freedom of the massive $U(1)$ vector. On the 
other hand, D~terms can provide positive-definite contributions to the tachyonic 
squared masses and, if allowed by flux quantization and by the full set of BI, they may 
even push these masses to positive values. This may happen also in the case of 
non-supersymmetric $AdS_4$ vacua, but it is more easily discussed in the case of 
supersymmetric $AdS_4$ vacua. For these solutions, in fact, D~terms vanish on 
the vacuum, because of supersymmetry, and the D-term contribution to the mass 
matrix of the geometrical moduli can be easily computed. In general, working in the
complex basis of the seven main moduli, and taking into account that the D-term 
potential depends only on the geometrical moduli and not on the axions, we can 
write for the normalized mass matrix:
\be
M^2_{kl} = M^2_{\ov k \ov l} = M^2_{k \ov l} = 
K^{1/2}_{\ov m k} K^{1/2}_{\ov l n} [(Re \, f)^{-1}]^{ab}
X_a^m \ov X_b^{\ov n} \, .
\ee
In the representative case under consideration, and around a 
vacuum with $\langle D \rangle = 0$, where $\langle s \rangle = s_0$ and 
$\langle u_A \rangle = u_{A \, 0}$ ($A=1,2,3$), with $q^0_a/s_0 =
\sum_{A=1}^3 (q^A_a / u_{A  \, 0})$, the D-term contribution to the
(normalized) mass matrix for the fields $(s,u_1,u_2,u_3)$ reads:
\beq
{\cal M}_{(D)}^2 = 
\frac{N \, T_6}{p_0 s_0 - \sum_{A=1}^3 p_A u_{A \, 0}}
\left(
\begin{array}{cc}
\frac{(q^0)^2}{s_0^2} \; & \; - \frac{q^0 \, q^A}{s_0 \, u_{A \, 0}} \\
&\\
- \frac{q^0 \, q^A}{s_0 \, u_{A \, 0}} \; & \; - \frac{q^A \, q^B}{u_{A \, 0} \, u_{B \, 0}}
\end{array}
\right) \, .
\eeq
The only non-vanishing eigenvalue is:
\beq
m_{(D)}^2 = \frac{N \, T_6}{p_0 s_0 - \sum_{A=1}^3 p_A u_{A \, 0}}
\left( 
\frac{(q^0)^2}{s_0^2} + \sum_{A=1}^3 \frac{(q^A)^2}{u_{A \, 0}^2} 
\right) \, ,
\eeq
and the corresponding eigenvector is
\beq
 \frac{q^0}{s_0^2} s - \sum_{A=1}^3 \frac{q^A}{u_{A \, 0}^2} u_A \, ,
\eeq
whose direction in the space of geometrical moduli is linked to the one of the 
gauged axions. We then see that, even though D~terms can be consistently 
neglected in the search for supersymmetric vacuum configurations, they do 
play an important r\^ole in the computation of the moduli spectrum.

We finally discuss how D~terms can remove some F-flat directions for the 
geometrical moduli in the case of Minkowski vacua. An important class
of these vacua is given by no-scale models, where supersymmetry is 
broken for all field configurations along one or more flat directions of the scalar
potential. The simplest way to obtain a no-scale model in the present context is
to introduce a superpotential that does not depend on three of the seven main
moduli, but depends non-trivially on the remaining four, so that the corresponding 
auxiliary fields can relax to zero on the vacuum. Using the index $\tilde k$ ($\hat k$) 
for the complex moduli that do (not) appear in the superpotential, we have in formulae:
$W_{\hat{k}} \equiv 0$, so that $G^{\hat k} G_{\hat k} \equiv K^{\hat k} K_{\hat k}  
\equiv 3$,  and $\langle G_{\tilde k} \rangle =0$. There are at least three geometrical 
moduli that are left to be stabilized, but at most three axionic shift symmetries that 
can be gauged and give rise to three independent D~terms: in the present class of
models, the only other axionic symmetry that could be gauged is broken by the 
superpotential. Since D~terms are 
homogeneous functions of the geometrical moduli, with three D~terms there is no way 
of stabilizing three geometrical moduli to non-zero values in a Minkowski background. 
With the inclusion of matter fields the analysis is more delicate since, in general, the 
assumption of complete factorizability of the K\"ahler manifold must be relaxed. 
However, for no-scale models realized with vanishing F~terms for the matter fields,  
the latter do not contribute to D~terms along the flat directions and the previous 
arguments still apply. We have checked on a number of examples that the above 
discussion seems to extend also to no-scale superpotentials with an explicit 
dependence on more than four out of the seven main moduli, once the various
consistency conditions are taken into account.

Another interesting class of Minkowski vacua are those where supersymmetry is 
preserved. As usual, $U(1)$ D~terms may be added if the corresponding axionic 
shift symmetries  are preserved by the superpotential. In a supersymmetric 
Minkowski background, since $ \langle W \rangle = 0$ and $ \langle W_i \rangle =0$, 
axionic shift symmetries get complexified \cite{complex}, so that the corresponding 
geometrical moduli are not stabilized by the F-term potential either. As discussed 
before, D~terms could just remove these flat directions of the scalar 
potential. However, even though we did not perform an exhaustive search, we  
are not aware of any explicit example, with all the consistency conditions satisfied, 
where all the main moduli are stabilized in this way. 

%%%%%%%%%%%%%%%%%%%%%%%%%%%%%%%%%%%%%%%
\section{Conclusions and outlook}

In this paper we studied the r\^ole of D~terms in string compactifications 
with D-branes and fluxes, preserving an exact or spontaneously broken
$\Ne1$ supersymmmetry in the effective field theory limit. In particular,
we neglected the matter field fluctuations,  and focused on the D-brane 
contributions to the potential for the closed string moduli, which we 
derived explicitly in the case of type-IIA O6 orientifold compactifications 
on the $T^6/(Z_2 \times Z_2)$ orbifold. We clarified under which 
assumptions the D-term breaking of supersymmetry can be considered 
soft, and how the effective D-brane actions can be embedded into the 
standard $\Ne1$ supergravity formalism. The D~terms depending on the 
closed string moduli are associated to the $U(1)$ gaugings that act as
shift symmetries on the RR axions. The latter can thus be absorbed by the 
brane vector fields, which become massive. The structure becomes more 
involved and more interesting when also bulk fluxes are considered. We 
showed that there is a strong connection between supersymmetry, gauge 
invariance, anomaly cancellation, Bianchi identities, flux superpotentials
and D~terms. The crucial point is the existence of constraints connecting 
D-branes and fluxes,  which guarantee the consistency of the effective theory 
both at the classical and at the quantum level. We derived these constraints
using localized Bianchi identities and T-duality, for $\Ne1$ compactifications
with generic RR, NSNS, geometrical and also non-geometrical fluxes.
These constraints dictate how D-branes can be embedded into a flux
compactification. The resulting effective action is automatically 
consistent with the gauging associated to the brane vector fields, 
and has vanishing gauge anomalies. Also, D terms appear to be
highly constrained by bulk fluxes, which in some cases are sufficient 
to completely specify their form. Our results are particularly relevant for 
string model building with flux compactifications, since they provide important
 constraints on these constructions.

We also discussed the r\^ole of D~terms for the problem of moduli stabilization.
As shown in the text, the D-term contribution to the moduli spectrum is always 
non-vanishing in all string compactifications where the chiral fields arise from 
intersecting (or magnetized) D-branes, even in the supersymmetric cases 
where D~terms vanish on the vacuum. We pointed out that, although
D~terms help removing residual flat directions of the F-term scalar potential,
in the chosen context there are some difficulties in reaching full stabilization 
of the closed string moduli in Minkowski or de-Sitter vacua. A systematic study 
of this aspect, however, goes beyond the purpose of the present paper.

There are various directions along which the results of the present paper
could be further developed.

An important step would be the incorporation of the chiral superfields,
corresponding to open string fluctuations living on branes or at brane
intersections, into the effective $\Ne1$ supergravity. In particular, fields
leaving at brane intersections are charged with respect to the gauged
$U(1)$s: in the present paper they have been neglected, but a more 
detailed analysis of their possible impact is certainly needed,
especially if we want to study supersymmetry-breaking vacua with
non-vanishing vacuum expectation values for both F and D~terms. Some
results on the K\"ahler potential for these "matter" fields are already
available \cite{lmrs}. It would be interesting to explore systematically, 
taking into account all available consistency constraints, whether the 
inclusion of matter fields can lead to new possibilities that do not seem to be
realized in their absence: for example, full moduli stabilization in Minkowski 
or de-Sitter vacua.

Last but not least, it would be interesting to include the effects of
warping in the derivation of the effective supergravity, and see whether
they can introduce any qualitatively new ingredients in the discussion of
supersymmetry breaking, moduli stabilization and the generation of
hierarchies.

 \vspace*{0.2cm}
 \noindent
{\bf Note added.} When the present paper was ready for submission, a new 
paper appeared \cite{acfi}, whose results 
partially overlap with our results on the effective
superpotentials with non-geometrical fluxes of sections 3.1 and 4. 
\acknowledgments
We thank A.~Uranga for discussions. This work was supported in part 
by the European Programme ``The Quest For Unification'', contract 
MRTN-CT-2004-503369.
%
%%%%%%%%%%%%%%%%%%%%%%%%%%%%%%%%%%%%%%%
\appendix
\section{Explicit results for the type-IIA $T^6/(Z_2 \times Z_2)$ orbifold with O6-planes}

We recall here the main formulae of sections~2 and~3, valid for flux compactifications 
of the type-IIA theory on the $T^6/(Z_2 \times Z_2)$ orbifold and O6 orientifold, in a more
explicit notation, so that they can be readily and unambiguously used for model-building.

The general form for the effective flux superpotential in $\Ne1$ type-IIA compactifications 
was given in eq.~(\ref{eq:superpot}). In components, and for the class of models under
consideration, we can also write:
\bea
4 \, W&=&
\Bigl ( \omega_{810}^{\quad 6} \, T_1 \, U_1+\omega_{106}^{\quad 8} 
\, T_2 \, U_2+\omega_{68}^{\quad 10} \, T_3 \, U_3\Bigr)
- S \, \Bigl (\omega_{79}^{\quad 6} \, T_1 + 
\omega_{95}^{\quad 8} \, T_2 +\omega_{57}^{\quad 10} \, T_3 \Bigr)
 \nn \\
% & & \nn \\ 
&& -\Bigl ( \omega_{89}^{\quad 5} \, T_1 \, U_3+
\omega_{96}^{\quad 7} \, T_2 \, U_3 + \omega_{710}^{\quad 5} 
\, T_1 \, U_2+\omega_{67}^{\quad 9} \, T_3 \, U_2
+\omega_{105}^{\quad 7} \, T_2 \, U_1+
\omega_{58}^{\quad 9} \, T_3 \, U_1\Bigr)\nn \\
% && \nn \\
&&
+ i \Bigl ( \ov G^{(4)}_{78910} \, T_1
+\ov G^{(4)}_{91056} \, T_2+\ov G^{(4)}_{5678}\, T_3\Bigr)-
\Bigl (\ov G^{(2)}_{56} \, T_2 \, T_3+\ov G^{(2)}_{78}\,  T_1 \, T_3
+\ov G^{(2)}_{910} \, T_1 \, T_2 \Bigr) \nn \\
% && \nn \\
&&
+ \ov G^{(6)} - i \, \ov G^{(0)} \, T_1 \, T_2 \, T_3
+i \Bigl ( \ov H_{579} \, S - \ov H_{5810} \, U_1
- \ov H_{6710} \, U_2- \ov H_{689} \, U_3 \Bigr)  \, .
\label{eq:explw}
\eea
Fluxes are constrained by the bulk and localized BI, which in the specific orbifold and 
orientifold under consideration give the following non-trivial relations:
\bea
&\omega_{mn}^{\quad \ p} \, \omega_{pr}^{\quad s}=0\,,& \label{eq:BIomega2} \\ &&\nn\\
&\l\{
\barr{ccc}
\frac12 \l ( \omega \, \ov G^{(2)}+\ov H \, \ov G^{(0)}\r)_{579}&=&\frac12\, \sum_a N_a\, p_{a \, 0} -16\,, \\
\frac12 \l ( \omega \, \ov G^{(2)}+\ov H \, \ov G^{(0)}\r)_{5810}&=&\frac12\, \sum_a N_a\, p_{a \, 1} +16\,, \\
\frac12 \l ( \omega \, \ov G^{(2)}+\ov H \, \ov G^{(0)}\r)_{6710}&=&\frac12\, \sum_a N_a\, p_{a \, 2} +16\,, \\
\frac12 \l ( \omega \, \ov G^{(2)}+\ov H \, \ov G^{(0)}\r)_{689}&=&\frac12\, \sum_a N_a\, p_{a \, 3} +16\,, 
\earr  \r.& \label{eq:BId6o6} \\ && \nn \\&&\nn\\
&\l\{ 
\barr{ccc}
\ov H_{579}\; q^0_a+\ov H_{5810}\; q^1_a+\ov H_{6710}\; q^2_a+\ov H_{689}\; q^3_a&=&0\,, \\
-\omega_{79}^{\quad 6}\, q^0_a -\omega_{810}^{\quad 6}\, q^1_a + \omega_{710}^{\quad 5}\, q^2_a+ \omega_{89}^{\quad 5}\, q^3_a&=&0\,,\\
-\omega_{95}^{\quad 8}\, q^0_a +\omega_{105}^{\quad 7}\, q^1_a - \omega_{106}^{\quad 8}\, q^2_a+ \omega_{96}^{\quad 7}\, q^3_a&=&0\,,\\
-\omega_{57}^{\quad 10}\, q^0_a+\omega_{58}^{\quad 9}\, q^1_a + \omega_{67}^{\quad 9}\, q^2_a- \omega_{68}^{\quad 10}\, q^3_a&=&0\,.
\earr \r.& \label{eq:BIloc}
\eea
where the sums run over the stacks of D6-branes and their images.
Besides the K\"ahler potential and the superpotential of eqs.~(\ref{kahler}) and (\ref{eq:explw}),
the remaining  ingredients needed to reconstruct the scalar potential of eq.~(\ref{vgen}) are
\be
\Re f_{ab} = \delta_{ab} N_a \, T_a \ \l(p_{a\,0} \, s- p_{a\,1} \, u_1-p_{a\,2} \, u_2-p_{a\,3} \, u_3 \r) \,,
\ee
\be
D_a = N_a \, T_a \l( \frac{q_a^0}{s}-\frac{q_a^1}{u_1}-\frac{q_a^2}{u_2}-\frac{q_a^3}{u_3}\r)\,,
\ee
and the supergravity description is valid as long as eq.~(\ref{eq:Dcond}) is satisfied,
\beq \label{eq:boundD}
\sqrt{s u_1 u_2 u_3} \l( \frac{q_a^0}{s}-\frac{q_a^1}{u_1}-\frac{q_a^2}{u_2}-\frac{q_a^3}{u_3}\r)\ll
\l(p_{a\,0}s-p_{a\,1}u_1-p_{a\,2}u_2-p_{a\,3}u_3 \r) \, .
\eeq
When this condition is not satisfied (large angles/D~terms, anti D-branes, \ldots), the 
scale of supersymmetry breaking lies outside the range of validity of the effective field
theory, which cannot be described anymore by the supergravity formalism above.

%%%%%%%%%%%%%%%%%%%%%%%%%%%%%%%%%%%%%%%%%%%%%%%%%%
%%                       BIBLIOGRAPHY                             %%
%%%%%%%%%%%%%%%%%%%%%%%%%%%%%%%%%%%%%%%%%%%%%%%%%%

%
\end{document}